\documentclass[aps,prl,twocolumn, 10pt, superscriptaddress]{revtex4-2}
\usepackage{graphicx}
\usepackage{amsmath}
\usepackage{amssymb}
\usepackage{mathtools}
\usepackage{braket}
\usepackage{xcolor}
\usepackage{physics}
\usepackage[colorlinks=true, citecolor=blue, urlcolor=blue, linkcolor=blue,bookmarks=false,hypertexnames=true]{hyperref} 
\usepackage[T1]{fontenc}

\usepackage{soul}
\renewcommand{\vec}[1]{\mathbf{#1}} 

\begin{document}
\graphicspath{}

\title{Bilayer graphene quantum dots as a quantum simulator of Haldane topological quantum matter}

\author{Daniel Miravet}
\thanks{dmiravet@uottawa.ca}
\affiliation{Department of Physics, University of Ottawa, Ottawa, Ontario, K1N 6N5, Canada}

\author{Hassan Allami}
\affiliation{Department of Physics, University of Ottawa, Ottawa, Ontario, K1N 6N5, Canada}

\author{Marek Korkusi\'nski}
\affiliation{Department of Physics, University of Ottawa, Ottawa, Ontario, K1N 6N5, Canada}
\affiliation{Quantum and Nanotechnologies Research Centre,
National Research Council of Canada, Ottawa, ON, K1A 0R6, Canada}

\author{Pawe\l\ Hawrylak}
\affiliation{Department of Physics, University of Ottawa, Ottawa, Ontario, K1N 6N5, Canada}

\date{\today}

\begin{abstract}
We demonstrate here that a chain of Bilayer Graphene Quantum Dots (BLGQDs)  can realize topological quantum matter by effectively simulating a spin-1 chain that hosts the Haldane phase within a specific range of parameters.
We describe a chain of BLGQDs with two electrons per dot using an atomistic tight-binding model combined with exact diagonalization to solve the interacting few-electron problem. Coulomb interactions and valley-mixing effects are treated within a single microscopic framework, allowing us to systematically investigate spin and valley polarization transitions as functions of interaction strength and external tuning parameters. We calculate the low energy states for single and double QDs as a function of the number of electrons, identifying regimes of highly correlated multi-electron states. We confirm the presence of a spin-one ground state for two electrons. Then, we explore two coupled QDs with 4 electrons and extend the analysis to QD arrays. Using a mapping of the BLGQD chain to an effective bilinear-biquadratic (BLBQ) spin model, we demonstrate that BLGQD arrays can work as a quantum simulator for one-dimensional spin chains with emergent many-body topological phases. 

\end{abstract}

\maketitle

Advancing quantum information technologies and uncovering novel topological phases of matter are central goals in modern condensed matter physics and quantum engineering~\cite{deep_am_2022,loss_arcmp_2013,korkusinski_worldscientific_2008, northup_nature_2014}.  A primary focus is the development of robust qubits with extended coherence times, with various platforms, including superconducting~\cite{sc_qb_anneal_nature_2011,top_transmon_2011,sc_qb_nature_2017,google_quantum_sup}, trapped ion~\cite{ion_trap_prx_2021,ion_trap_read_out_2008,ion_trap_wright2019}, photonic~\cite{loqc_2001,loqc_ph_qb_2007,xanadu_quantum_advantage,northup2014quantum}, and semiconductor spin qubits~\cite{coupled_qd_qb_1997,qc_qd_1997,korkusinski_worldscientific_2008,nano_qi_2023,spin_qd_qubit_2013,so_qubit_si_2021,coherent_spin_qd_2006,store_spin_si_2014,Petta_STQB_2005,Ciorga2000,GaudreauPRL2006,PioroPRL2021}.  Spin qubits and circuits, particularly spin chains, stand out as prototypes for investigating topologically strongly correlated quantum matter, particularly due to their capacity to host  Haldane spin-1/2 quasiparticles at their edges~\cite{haldane_physlett_1983, haldane_rev_mod_phys_2017, affleck_prl_1987,jaworowski_scirep_2017}.

The artificial synthetic spin-1 chains have been intensely studied, but their experimental realization is limited. Spin-1 chains can be engineered using various quantum dot architectures, such as gated triple quantum dots \cite{shim-hawrylak-ssc2010}, linear arrays of semiconductor quantum dots embedded in nanowires \cite{lafferiere_apl_2021,jaworowski_scirep_2017, manalo_prb_2021, AllamiPRBTwoQubits}, chains formed by triangular graphene quantum dots \cite{devrim_springer_2014, fasel_nature_2021, rossier_prb_2022,YasserSuperexchangeNanoletter2024}, or tunable hybrid platform of superconducting islands and quantum dots~\cite{Virgilsuperconductor-semiconductorPRB2024}. For instance, chains of InAsP quantum dots embedded in an InP nanowire have been proposed and predicted to host such synthetic spin-1 objects~\cite{manalo_prb_2021,ManaloPRBSpin1chain}, realizing macroscopic quantum states in semiconductors. In these systems, four electrons per quantum dot can form a synthetic spin-1 state, and their low-energy behaviour can be effectively described by a Hubbard-Kanamori Hamiltonian derived from atomistic microscopic calculations~\cite{ManaloPRBSpin1chain}. This framework has successfully demonstrated that these arrays can emulate antiferromagnetic spin-1 chains, providing a foundation for engineering synthetic topologically nontrivial quantum matter. 
Gated bilayer graphene (BLG) quantum dots (QDs)~\cite{pereira2007tunable,recherGrapheneQD2009} offer a particularly attractive platform for this purpose. Single and double BLGQDs have been experimentally realized, with evidence of spin‑1 ground states in few‑electron configurations~\cite{MollerPRL2021,korkusinski2023spontaneous}. BLG is unique among two‑dimensional materials in that a perpendicular electric field can open a tunable band gap, enabling electrical control over confinement and access to both spin and valley degrees of freedom~\cite{MollerPRL2021,Matthew_PRB_TG_2024,korkusinski2023spontaneous}. Recent theoretical and experimental studies of BLGQDs show that two confined electrons can robustly form a triplet ground state across a wide range of interaction strengths, often accompanied by spontaneous spin–valley polarization~\cite{korkusinski2023spontaneous,MollerPRL2021}. This robust triplet formation makes BLGQDs promising building blocks for electrically tunable synthetic spin‑1 chains.

Here, we investigate single and coupled BLGQDs using an atomistic tight‑binding model combined with exact diagonalization to treat Coulomb interactions and valley mixing on equal footing. For two coupled QDs, we show that the low‑energy sector maps naturally onto a bilinear–biquadratic spin‑1 model, establishing a direct link between microscopic BLGQD states and effective spin‑1 chains. This mapping highlights BLGQDs as a practical and tunable platform for exploring one-dimensional quantum magnetism, topological phases, and electrically controlled quantum information architectures.



\begin{figure}[ht]
	\includegraphics[width=\columnwidth]{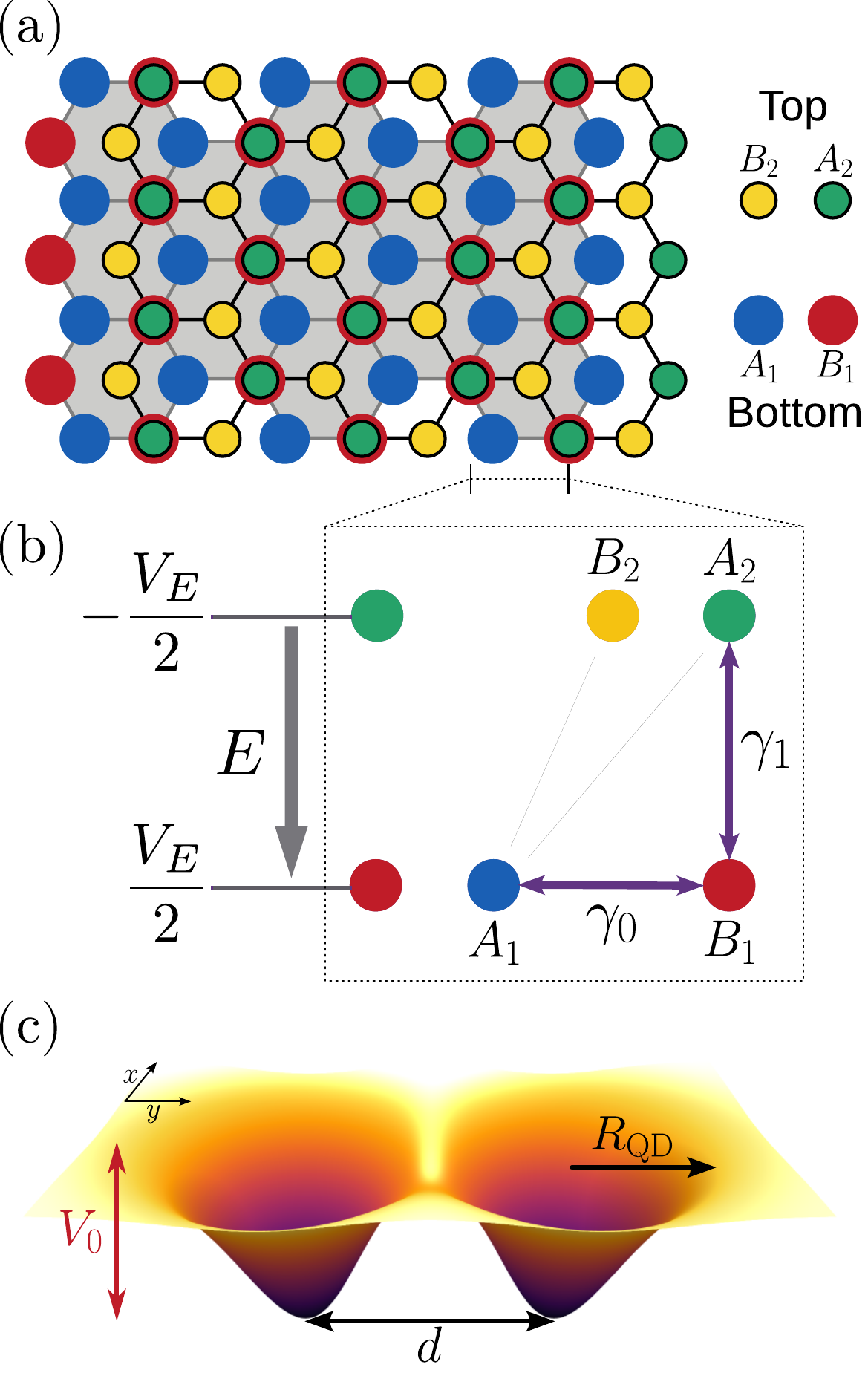}
	\caption{(a) Top view of bilayer graphene, showing the four atoms within the unit cell:  $A_1$, $B_1$, $A_2$, $B_2$, distinguished by colour.
    (b) Side view (zoomed in), highlighting the two primary hopping parameters: $\gamma_0$ for intralayer coupling and $\gamma_1$ for interlayer coupling. The applied perpendicular electric field is shown schematically; note that the diagram depicts the resulting potential energy profile rather than the electrostatic potential.
    (c) Schematic illustration of the confining potential for a double quantum dot. The parameters $V_0$, $R_{\rm{QD}}$, and $d$ denote the potential depth, dot radius, and center-to-center distance between the dots, respectively.
    }
	\label{fig:FigOnePNG}
\end{figure}

We model Bernal‑stacked bilayer graphene (BLG), Fig.~\ref{fig:FigOnePNG}(a), whose bottom (top) layer contains sublattices $A_1$ and $B_1$ ($A_2$ and $B_2$). The electronic structure is described within a tight‑binding model including the dominant in‑plane and interlayer hopping parameters ($\gamma_0$, $\gamma_1$) and a perpendicular electric field that opens a tunable band gap [Fig.~\ref{fig:FigOnePNG}(b)].

Gate‑defined quantum dots (QDs) are created by adding smooth Gaussian confinement potentials [Fig.~\ref{fig:FigOnePNG}(c)], allowing us to realize single‑ or double‑dot geometries. To obtain the confined single-particle states, we expand the full Hamiltonian $ \hat{H}_{\rm{QD}} = \hat{H}_{\textrm{bulk}} + \hat{V}_{\rm{ext}}$ in a basis of bulk Bloch states near the $\pm K$ valleys and diagonalize it in momentum space. This approach captures the effect of the smooth confinement without resorting to large-scale real-space atomistic calculations \cite{Miravet_PRB_QD_Holes}.

Once the confined orbitals are obtained, we compute all relevant Coulomb matrix elements between them and perform exact diagonalization of the many-body Hamiltonian to access the interacting few-electron spectrum. Numerically, we sample the Brillouin zone on a rhomboidal supercell of $901\times 901$ k-points (equivalent to a real-space system of $\sim 3\times10^6$ atoms), ensuring convergence of both the single-particle and interacting energy levels. Full details of the tight-binding model, the momentum-space confinement treatment, and the numerical implementation are provided in the Supplementary Material.

Many‑body effects are incorporated by filling the lowest single‑particle levels with $N$ electrons and adding the Coulomb interaction,

\begin{equation}
H=\sum_{s} E_s c^\dagger_s c_s +  \frac{\lambda}{2}\sum_{p,q,r,s} \bra{pq}V\ket{rs} c^\dagger_p c^\dagger_q c_r c_s.
\label{eq:manybodyHam}
\end{equation}
where $E_s$ are the single‑particle energies, $\langle pq|V|rs\rangle$ are the Coulomb matrix elements, and $\lambda\in[0,1]$ tunes the interaction strength. We model electron-electron interactions using a scaled Coulomb potential with a dimensionless parameter $\lambda$ that captures the effect of dielectric screening and gate geometry. The parameter $\lambda$ effectively renormalizes the interaction strength, allowing us to explore different interaction regimes. In experimental BLGQD systems, $\lambda$ can be tuned via the dielectric environment (e.g., hBN thickness and number of layers), the distance to metallic gates, and the quantum dot size. The spin structure is included explicitly, and all matrix elements are evaluated using the numerically obtained single‑particle orbitals. Interdot electron-electron interactions are naturally included through the long-range Coulomb interaction, as we compute all matrix elements between all confined orbitals in the system.

The many-body Hamiltonian in the space of multi-electron configurations is solved via exact diagonalization to obtain the low-energy spectrum~\cite{korkusinski2023spontaneous, Miravet_PRB_QD_Holes}. We perform the diagonalization in the full basis of confined orbitals, imposing only particle-number conservation as a constraint. At the single-particle level, the smooth confinement preserves valley symmetry (lacking Fourier components at $|K-K'|$), so Bloch states near $\pm K$ decouple and the numbers of $K$ and $K'$ states are conserved in the non-interacting problem. For the Coulomb interaction, we compute all matrix elements between all confined orbitals, including both intravalley and intervalley direct and exchange terms.  Momentum-transfer selection rules make intervalley Coulomb scattering small but nonzero, allowing for valley mixing in the many-body states. Despite not explicitly imposing spin or valley conservation in the many-body calculation, the resulting eigenstates naturally organize according to total valley index, total spin $S$, and $S_z$, reflecting the underlying symmetries of the Hamiltonian.

For completeness, we first characterize the single quantum dot (QD) spectrum as a reference for the analysis of the double QD and BLG QD chain. For a representative set of parameters, the single-particle levels form valley doublets, which become fourfold degenerate when spin is included, and display shell structures reminiscent of a two-dimensional harmonic oscillator with additional lattice-induced splittings. Including Coulomb interactions for two electrons, we find that the ground state remains a triplet over the entire interaction range considered, with a crossover in valley polarization from polarized to unpolarized as the interaction weakens. The corresponding low-energy spectrum, along with its classification by total spin and valley polarization, is presented in the Supplementary Material, which also includes full computational details and analytical considerations.
 
 Motivated by the observation that the $N=2$ ground state remains a triplet across a broad range of interaction strengths, we next examine the case of a double quantum dot (dQD) populated by four electrons. We aim to compare the resulting low-energy spectrum with that of an effective spin-1 Heisenberg model. To this end, we begin by analyzing the single-particle states of the double quantum dot, and then determine the interacting four-electron ground state along with its low-lying excitations.



\begin{figure}[ht]
\centering     
\includegraphics[width=\columnwidth]{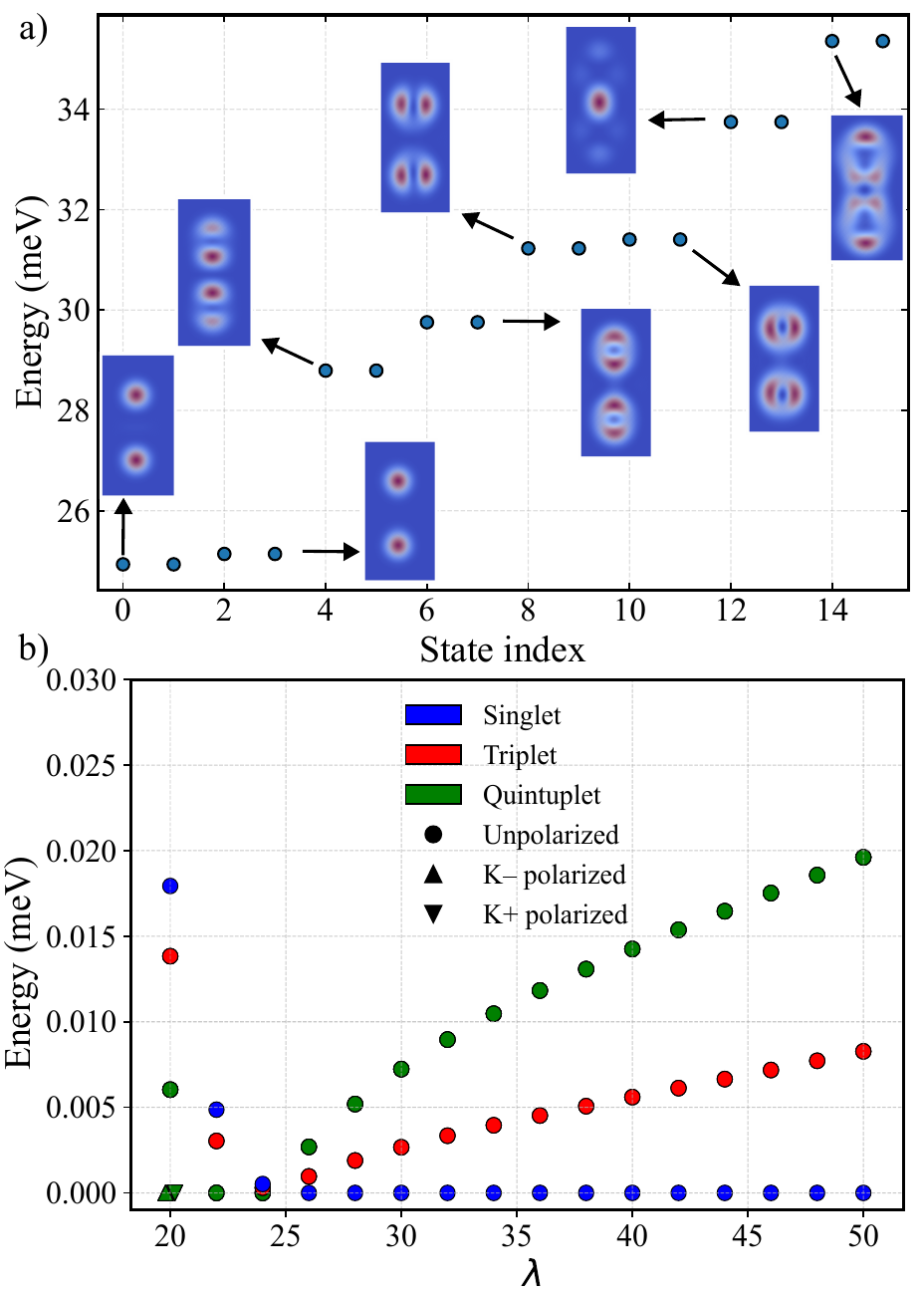}
\caption{(a) Low-energy levels of a double quantum dot. Each state shown is doubly degenerate by spin; only the valley-degenerate states are explicitly labeled. Insets show the corresponding electron densities for each shell, illustrating the increased coupling between interdot states with higher energy, as a consequence of the greater spatial extension of the wavefunctions.  
(b) Low-energy spectrum for four electrons in a double quantum dot as a function of interaction strength. For weak interactions, the ground state is a valley-unpolarized singlet, while for strong interactions, it becomes a valley-polarized quintuplet. There is an artificial horizontal shift for the valley-polarized quintuplets to show both valley cases. In the regime where the ground state is a singlet, the first and second excited states are a triplet and a quintuplet, respectively, closely resembling the spectrum of a spin-1 Heisenberg antiferromagnetic chain with $L = 2$ sites.}

\label{fig:DQD_energy_levels}
\end{figure}

Figure~\ref{fig:DQD_energy_levels} shows the single-particle spectrum of the double quantum dot for an interdot separation $d=40$ nm. Compared to the single QD case, each shell is now duplicated, with an energy splitting which reflects the presence of another coupled dot. This behaviour can be understood by starting from the single-dot spectrum: as the dots are brought closer together, hybridization between the individual QD states increases, leading to the formation of bonding and antibonding combinations. Consequently, the splitting between these paired states grows as the interdot distance decreases and the depth of the potential increases~\cite{dQDPeeters2017,PawlowskiPRA2021,dQDKnothePRB2024}. 

Since higher QD states have a more extended wave function, the state hybridization is stronger for higher energy shells. We can see it on the probability density plot for each state (Insets in Fig.~\ref{fig:DQD_energy_levels}). Low-energy states like the one in the s-shell are not strongly modified compared to the single QD version. On the other hand, states composed of QD p and d shells are more modified, with the effect being bigger as the state energy increases. 

Figure~\ref{fig:DQD_energy_levels}(b) shows the low-energy spectrum for four electrons confined in a double quantum dot as a function of interaction strength. In the weak-interaction regime, the ground state remains a valley-unpolarized singlet, with the first and second excited states being a triplet and a quintuplet, respectively. The main contribution to the singlet state comes from the configuration with the lowest shell completely full. This sequence closely matches the spectrum of a spin-1 Heisenberg antiferromagnetic chain with $L = 2$ sites, consistent with the mapping to an effective spin model.

As interaction strength increases, the system transitions to a valley-polarized quintuplet ground state ($S=2$). This transition is driven by electron-electron repulsion and fine tuned by the exchange interaction: strong interactions make it energetically favorable to promote electrons to the second shell (at a kinetic energy cost), and since intravalley exchange is significantly larger than intervalley exchange, the electrons tend to concentrate in a single valley. In this case the next excited states are a quintuplet, followed by a triplet and then a singlet, mainly composed of electrons on the lowest shell. Remarkably, at intermediate interaction, when the ground state becomes a valley-unpolarized quintuplet, the first and second excited states are a triplet and a singlet, respectively, mirroring the spectrum expected for two spin-1s coupled ferromagnetically.

\begin{figure}[ht]
\centering     
\includegraphics[width=\columnwidth]{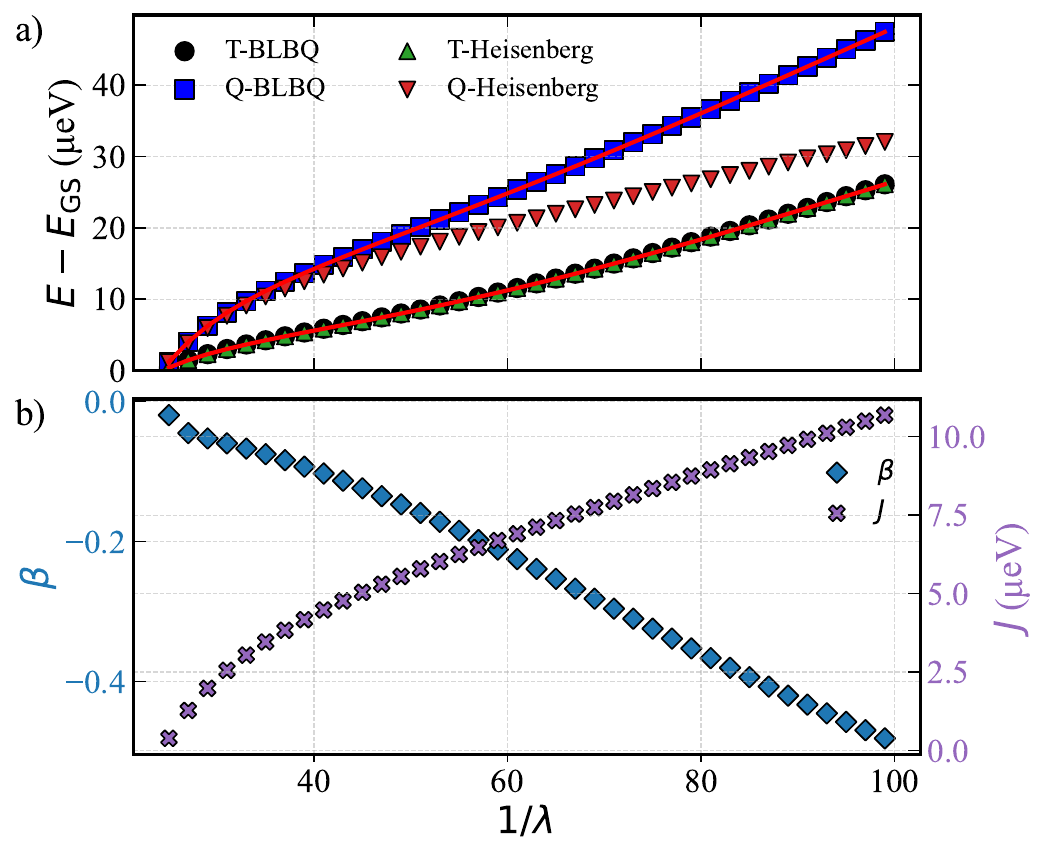}

\caption{
(a) Triplet and quintuplet energies obtained from the fitted BLBQ and Heisenberg models as a function of interaction strength. The blue dashed lines show the energies of the double quantum dot. As discussed in the main text, the BLBQ Hamiltonian accurately reproduces the low-energy spectrum of two coupled quantum dots, each containing two electrons.  
(b) Fitted values of the BLBQ parameters $\beta$ and $J$ as functions of the electron-electron interaction strength.}
  
\label{fig:BLBQfit}
\end{figure}

The bilinear–biquadratic  (BLBQ) spin-1 Hamiltonian 
\begin{equation}
H=J\sum_{i}^{L-1} \left( \vec{S}_i\cdot  \vec{S}_{i+1} + \beta \left( \vec{S}_i\cdot  \vec{S}_{i+1} \right)^2 \right)
\label{eq:manybodyHam___}
\end{equation}
where $\vec{S}_i$ is the spin operator acting on site $i$, and $J$ is the exchange coupling, $\beta$ is a tunable parameter that controls the strength of the biquadratic interaction, and $L$ is the number of sites. For even $L$, and $\beta < 1/3$, the low-energy spectrum consists of a singlet ground state, followed by a triplet and a quintuplet. In particular, for two spins, the energy gaps are given by (see SM for details), $\Delta_{TQ}=2J$, $\Delta_{ST}=J(1-3\beta)$. The bilinear-biquadratic spin-1 model exhibits the Haldane phase for $-1 < \beta < 1$, a topologically ordered phase first predicted by Haldane for the $S=1$ antiferromagnetic Heisenberg chain. This phase is characterized by a gapped bulk spectrum and fractionalized spin-1/2 edge states, despite the bulk being made of integer spins~\cite{haldane_physlett_1983,affleck_prl_1987}.

\begin{figure}[ht]
\centering     
\includegraphics[width=\columnwidth]{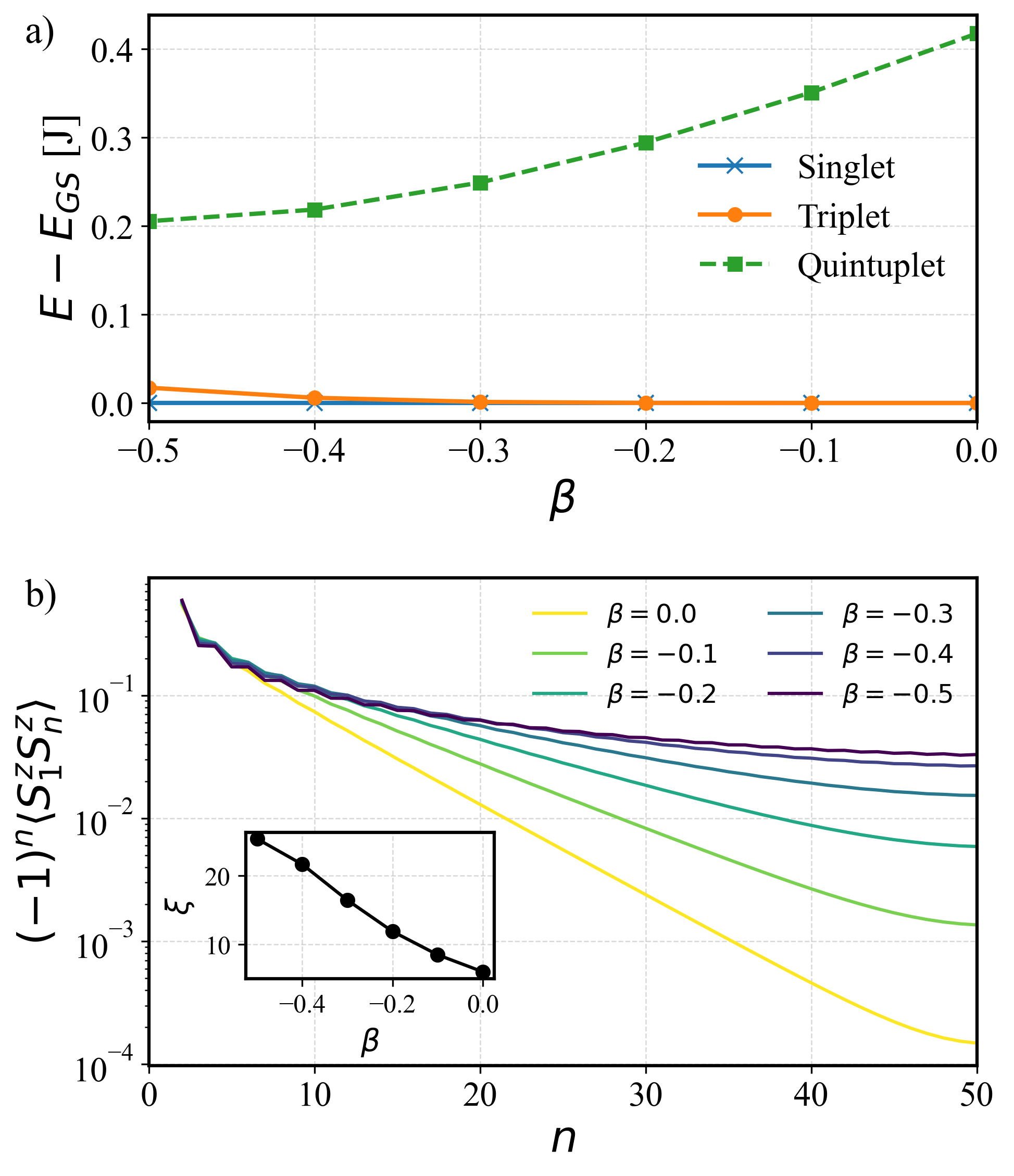}

\caption{
(a) Low energies of the $L=100$ BLBQ spin chain as a function of the  $\beta$.  
(b) Spin-spin correlation function $g(n) = (-1)^n \langle S_1^z S_n^z \rangle$ plotted for the first half of the chain. The inset shows the correlation length $\xi$ extracted from exponential fits to $g(n)$, illustrating its dependence on $\beta$. The correlation length $\xi$ characterizes the spatial decay of spin correlations in the system.}

\label{fig:BLBQ_plots}
\end{figure}

Using these expressions, we can map the dQD spectrum onto an effective BLBQ model and extract the corresponding $J$ and $\beta$ parameters. Figure~\ref{fig:BLBQfit}(a) shows the low energy levels from the fitted BLBQ and Heisenberg models as a function of interaction strength. The blue dashed lines show the energies of the double quantum dot. The BLBQ Hamiltonian reproduces exactly the two low-energy gaps of two coupled quantum dots, containing two electrons each. Figure~\ref{fig:BLBQfit}(b) shows the fitted value of $J$ and $\beta$ for different values of the ratio between kinetic and Coulomb interacting terms. 

For stronger interactions, $\beta$ is near zero, becoming more negative as the interaction strength decreases. This behavior is consistent with numerical~\cite{soniBubbard2BLBQ2022} and perturbative~\cite{tanakaOriginBiquadraticExchange2018} results in multiorbital Hubbard models, where the largest  values of $|\beta|$ are obtained when the Hubbard interaction $U$ is reduced. In the perturbative limit where an expansion in $t/U$ is valid, the bilinear term $J$ appears at lowest order while the biquadratic term (proportional to $\beta$) appears at next-to-leading order. The minimum value shown here for weak interaction is $\beta\approx-0.5$, which is still inside the Haldane phase region. The sign of $\beta$ reflects the structure of the effective spin Hamiltonian emerging from multiorbital superexchange~\cite{tanakaOriginBiquadraticExchange2018,soniBubbard2BLBQ2022}.

To explore the behavior of the BLBQ Hamiltonian, we computed the low-energy spectrum of a spin-1 chain with $L = 100$ using matrix product state–based density matrix renormalization group (MPS-DMRG) approach~\cite{schollwock_annals_of_physics_2011,schollwock_rev_mod_phys_2005,white_prl_1992}. Figure~\ref{fig:BLBQ_plots}(a) shows the low-energy spectrum as a function of the biquadratic interaction parameter $\beta$ in the range $-1/2 \leq \beta \leq 0$. In the entire Haldane phase ($-1 < \beta < 1$) and in the thermodynamic limit, the ground state is fourfold degenerate, consisting of a singlet and a triplet manifold separated from the lowest quintuplet by a finite topological gap. This degeneracy arises from two fractionalized spin-$\tfrac{1}{2}$ quasiparticles localized at the chain ends. In a finite chain, the singlet–triplet gap is nonzero but decreases exponentially with system size, reflecting the exponentially small coupling between the edge spins. In contrast, the topological gap to the quintuplet remains finite in the thermodynamic  limit~\cite{haldane_physlett_1983,affleck_prl_1987}.

Figure~\ref{fig:BLBQ_plots}(b) displays the spin-spin correlation function $g(n) = (-1)^n \left\langle S_1^z S_n^z \right\rangle$ for the first half of the chain. As $\beta$ becomes more negative, the decay of $g(n)$ slows, indicating an increase in the correlation length $\xi$. This length scale is extracted by fitting the envelope of the data to an exponential form, $g(n) \propto \exp(-n/\xi)$. The inset shows the fitted values of $\xi$ as a function of $\beta$. The correlation length $\xi$ controls both the spatial extent of the edge quasiparticles and the range over which they can interact. When $L \gg \xi$, the system is effectively in the thermodynamic limit and the singlet–triplet splitting becomes negligible. However, for $\beta \approx -0.5$, $\xi$ becomes comparable to $L = 100$, making finite-size effects visible in the spectrum and enhancing the singlet–triplet splitting. As $\beta \to -1$, the system approaches a quantum critical point characterized by a vanishing triplet-quintuplet gap and a diverging correlation length~\cite{RakovBLBQHaldanePRB2022}. In this regime, DMRG convergence becomes increasingly challenging due to enhanced entanglement and critical fluctuations.

The mapping from BLGQDs to the BLBQ model can be understood within a broader theoretical framework. Similar to the two-site two-orbital Hubbard model~\cite{soniBubbard2BLBQ2022,tanakaOriginBiquadraticExchange2018}, our graphene QD with two electrons exhibits a half-filled degenerate $s$-shell at low energies, where the two orbital levels play the role of the two orbitals per site in the Hubbard model. Under the effective condition of weaker interactions, both systems map to the same BLBQ Hamiltonian, demonstrating the universality of this effective description for systems with degenerate orbital manifolds.

The correlation length $\xi$ determines the minimal chain length required to observe 
topological properties. For the Haldane phase with $\beta$ near 0, $\xi \sim 6$ sites, 
so chains of length $L \approx 10-20$ sites should be sufficient to observe clear 
topological signatures such as the bulk gap and edge states. For example, topological 
effects have been observed in chains of 16 nanographene quantum dots~\cite{fasel_nature_2021}. 
Moreover, topological properties such as edge states and bulk gap are observable even 
in relatively short chains, making BLGQDs a promising platform for quantum simulation 
despite current experimental constraints.

The robustness of our results against model details has been verified by including the skew-hoppings $\gamma_3$ and $\gamma_4$ (see Supplementary Material). These terms introduce only quantitative corrections (trigonal warping and particle–hole asymmetry) but do not affect the qualitative physics: the spin structure, ground state nature, and valley ordering of the low-energy manifold remain unchanged~\cite{sadecka2023electrically,Matthew_PRB_TG_2024}. This robustness stems from the fact that the spin-1 character is determined by the fourfold degenerate lowest shell, which is insensitive to these corrections.


In summary, we have demonstrated that chains of electrostatically defined bilayer graphene quantum dots (BLGQDs) can become synthetic spin-1 chains, providing an atomistic platform for topological quantum matter. Using a realistic tight‑binding description combined with exact diagonalization, we captured Coulomb interactions and valley mixing on equal footing for both single‑ and double‑dot configurations. For two coupled QDs, the low‑energy spectrum maps naturally onto a bilinear–biquadratic spin‑1 model, establishing a direct link between microscopic electronic states in BLGQDs and effective quantum spin Hamiltonians. This correspondence opens a route to quantum simulators of Haldane topological quantum matter, one‑dimensional quantum magnetism with electrically tunable parameters, offering new opportunities for quantum simulation, spintronics, and solid‑state quantum information processing.

\begin{acknowledgments}
The authors thank Christoph Stampfer and Hubert Dulisch for valuable discussions. This work was supported by the Quantum Sensors Challenge Program QSP-078, High Throughput Networks HTSN-341, and Applied Quantum Computing AQC-004 Challenge Programs at the National Research Council of Canada, NSERC Discovery Grant No. RGPIN-2019-05714,
NSERC PQS2D Alliance Quantum Grant No. ALLRP/578466-2022, and University of Ottawa Research Chair in Quantum Theory
of Materials, Nanostructures, and Devices.  This work was partly enabled by the support provided by the Digital Research
Alliance of Canada~\cite{alliancecan2025}.
\end{acknowledgments}

\bibliography{main}

\begin{thebibliography}{58}%
\makeatletter
\providecommand \@ifxundefined [1]{%
 \@ifx{#1\undefined}
}%
\providecommand \@ifnum [1]{%
 \ifnum #1\expandafter \@firstoftwo
 \else \expandafter \@secondoftwo
 \fi
}%
\providecommand \@ifx [1]{%
 \ifx #1\expandafter \@firstoftwo
 \else \expandafter \@secondoftwo
 \fi
}%
\providecommand \natexlab [1]{#1}%
\providecommand \enquote  [1]{``#1''}%
\providecommand \bibnamefont  [1]{#1}%
\providecommand \bibfnamefont [1]{#1}%
\providecommand \citenamefont [1]{#1}%
\providecommand \href@noop [0]{\@secondoftwo}%
\providecommand \href [0]{\begingroup \@sanitize@url \@href}%
\providecommand \@href[1]{\@@startlink{#1}\@@href}%
\providecommand \@@href[1]{\endgroup#1\@@endlink}%
\providecommand \@sanitize@url [0]{\catcode `\\12\catcode `\$12\catcode `\&12\catcode `\#12\catcode `\^12\catcode `\_12\catcode `\%12\relax}%
\providecommand \@@startlink[1]{}%
\providecommand \@@endlink[0]{}%
\providecommand \url  [0]{\begingroup\@sanitize@url \@url }%
\providecommand \@url [1]{\endgroup\@href {#1}{\urlprefix }}%
\providecommand \urlprefix  [0]{URL }%
\providecommand \Eprint [0]{\href }%
\providecommand \doibase [0]{https://doi.org/}%
\providecommand \selectlanguage [0]{\@gobble}%
\providecommand \bibinfo  [0]{\@secondoftwo}%
\providecommand \bibfield  [0]{\@secondoftwo}%
\providecommand \translation [1]{[#1]}%
\providecommand \BibitemOpen [0]{}%
\providecommand \bibitemStop [0]{}%
\providecommand \bibitemNoStop [0]{.\EOS\space}%
\providecommand \EOS [0]{\spacefactor3000\relax}%
\providecommand \BibitemShut  [1]{\csname bibitem#1\endcsname}%
\let\auto@bib@innerbib\@empty
\bibitem [{\citenamefont {Alfieri}\ \emph {et~al.}(2022)\citenamefont {Alfieri}, \citenamefont {Anantharaman}, \citenamefont {Zhang},\ and\ \citenamefont {Jariwala}}]{deep_am_2022}%
  \BibitemOpen
  \bibfield  {author} {\bibinfo {author} {\bibfnamefont {A.}~\bibnamefont {Alfieri}}, \bibinfo {author} {\bibfnamefont {S.~B.}\ \bibnamefont {Anantharaman}}, \bibinfo {author} {\bibfnamefont {H.}~\bibnamefont {Zhang}},\ and\ \bibinfo {author} {\bibfnamefont {D.}~\bibnamefont {Jariwala}},\ }\bibfield  {title} {\bibinfo {title} {Nanomaterials for quantum information science and engineering},\ }\href {https://doi.org/https://doi.org/10.1002/adma.202109621} {\bibfield  {journal} {\bibinfo  {journal} {Advanced Materials}\ ,\ \bibinfo {pages} {2109621}} (\bibinfo {year} {2022})}\BibitemShut {NoStop}%
\bibitem [{\citenamefont {Kloeffel}\ and\ \citenamefont {Loss}(2013{\natexlab{a}})}]{loss_arcmp_2013}%
  \BibitemOpen
  \bibfield  {author} {\bibinfo {author} {\bibfnamefont {C.}~\bibnamefont {Kloeffel}}\ and\ \bibinfo {author} {\bibfnamefont {D.}~\bibnamefont {Loss}},\ }\bibfield  {title} {\bibinfo {title} {Prospects for spin-based quantum computing in quantum dots},\ }\href {https://doi.org/10.1146/annurev-conmatphys-030212-184248} {\bibfield  {journal} {\bibinfo  {journal} {Annual Review of Condensed Matter Physics}\ }\textbf {\bibinfo {volume} {4}},\ \bibinfo {pages} {51} (\bibinfo {year} {2013}{\natexlab{a}})}\BibitemShut {NoStop}%
\bibitem [{\citenamefont {Korkusinski}\ and\ \citenamefont {Hawrylak}(2008)}]{korkusinski_worldscientific_2008}%
  \BibitemOpen
  \bibfield  {author} {\bibinfo {author} {\bibfnamefont {M.}~\bibnamefont {Korkusinski}}\ and\ \bibinfo {author} {\bibfnamefont {P.}~\bibnamefont {Hawrylak}},\ }\bibinfo {title} {Coded qubits based on electron spin, in {S}emiconductor {Q}uantum {B}its}\ (\bibinfo  {publisher} {Pan Stanford Publishing},\ \bibinfo {year} {2008})\ Chap.~\bibinfo {chapter} {1}, pp.\ \bibinfo {pages} {3--32}\BibitemShut {NoStop}%
\bibitem [{\citenamefont {Northup}\ and\ \citenamefont {Blatt}(2014{\natexlab{a}})}]{northup_nature_2014}%
  \BibitemOpen
  \bibfield  {author} {\bibinfo {author} {\bibfnamefont {T.~E.}\ \bibnamefont {Northup}}\ and\ \bibinfo {author} {\bibfnamefont {R.}~\bibnamefont {Blatt}},\ }\bibfield  {title} {\bibinfo {title} {Quantum information transfer using photons},\ }\href {https://doi.org/10.1038/nphoton.2014.53} {\bibfield  {journal} {\bibinfo  {journal} {Nature Photonics}\ }\textbf {\bibinfo {volume} {8}},\ \bibinfo {pages} {356} (\bibinfo {year} {2014}{\natexlab{a}})}\BibitemShut {NoStop}%
\bibitem [{\citenamefont {Johnson}\ \emph {et~al.}(2011)\citenamefont {Johnson}, \citenamefont {Amin}, \citenamefont {Gildert}, \citenamefont {Lanting}, \citenamefont {Hamze}, \citenamefont {Dickson}, \citenamefont {Harris}, \citenamefont {Berkley}, \citenamefont {Johansson}, \citenamefont {Bunyk}, \citenamefont {Chapple}, \citenamefont {Enderud}, \citenamefont {Hilton}, \citenamefont {Karimi}, \citenamefont {Ladizinsky}, \citenamefont {Ladizinsky}, \citenamefont {Oh}, \citenamefont {Perminov}, \citenamefont {Rich}, \citenamefont {Thom}, \citenamefont {Tolkacheva}, \citenamefont {Truncik}, \citenamefont {Uchaikin}, \citenamefont {Wang}, \citenamefont {Wilson},\ and\ \citenamefont {Rose}}]{sc_qb_anneal_nature_2011}%
  \BibitemOpen
  \bibfield  {author} {\bibinfo {author} {\bibfnamefont {M.~W.}\ \bibnamefont {Johnson}}, \bibinfo {author} {\bibfnamefont {M.~H.}\ \bibnamefont {Amin}}, \bibinfo {author} {\bibfnamefont {S.}~\bibnamefont {Gildert}}, \bibinfo {author} {\bibfnamefont {T.}~\bibnamefont {Lanting}}, \bibinfo {author} {\bibfnamefont {F.}~\bibnamefont {Hamze}}, \bibinfo {author} {\bibfnamefont {N.}~\bibnamefont {Dickson}}, \bibinfo {author} {\bibfnamefont {R.}~\bibnamefont {Harris}}, \bibinfo {author} {\bibfnamefont {A.~J.}\ \bibnamefont {Berkley}}, \bibinfo {author} {\bibfnamefont {J.}~\bibnamefont {Johansson}}, \bibinfo {author} {\bibfnamefont {P.}~\bibnamefont {Bunyk}}, \bibinfo {author} {\bibfnamefont {E.~M.}\ \bibnamefont {Chapple}}, \bibinfo {author} {\bibfnamefont {C.}~\bibnamefont {Enderud}}, \bibinfo {author} {\bibfnamefont {J.~P.}\ \bibnamefont {Hilton}}, \bibinfo {author} {\bibfnamefont {K.}~\bibnamefont {Karimi}}, \bibinfo {author} {\bibfnamefont {E.}~\bibnamefont {Ladizinsky}}, \bibinfo {author} {\bibfnamefont
  {N.}~\bibnamefont {Ladizinsky}}, \bibinfo {author} {\bibfnamefont {T.}~\bibnamefont {Oh}}, \bibinfo {author} {\bibfnamefont {I.}~\bibnamefont {Perminov}}, \bibinfo {author} {\bibfnamefont {C.}~\bibnamefont {Rich}}, \bibinfo {author} {\bibfnamefont {M.~C.}\ \bibnamefont {Thom}}, \bibinfo {author} {\bibfnamefont {E.}~\bibnamefont {Tolkacheva}}, \bibinfo {author} {\bibfnamefont {C.~J.~S.}\ \bibnamefont {Truncik}}, \bibinfo {author} {\bibfnamefont {S.}~\bibnamefont {Uchaikin}}, \bibinfo {author} {\bibfnamefont {J.}~\bibnamefont {Wang}}, \bibinfo {author} {\bibfnamefont {B.}~\bibnamefont {Wilson}},\ and\ \bibinfo {author} {\bibfnamefont {G.}~\bibnamefont {Rose}},\ }\bibfield  {title} {\bibinfo {title} {Quantum annealing with manufactured spins},\ }\href {https://doi.org/https://doi.org/10.1038/nature10012} {\bibfield  {journal} {\bibinfo  {journal} {Nature}\ }\textbf {\bibinfo {volume} {473}},\ \bibinfo {pages} {194} (\bibinfo {year} {2011})}\BibitemShut {NoStop}%
\bibitem [{\citenamefont {Hassler}\ \emph {et~al.}(2011)\citenamefont {Hassler}, \citenamefont {Akhmerov},\ and\ \citenamefont {Beenakker}}]{top_transmon_2011}%
  \BibitemOpen
  \bibfield  {author} {\bibinfo {author} {\bibfnamefont {F.}~\bibnamefont {Hassler}}, \bibinfo {author} {\bibfnamefont {A.~R.}\ \bibnamefont {Akhmerov}},\ and\ \bibinfo {author} {\bibfnamefont {C.~W.~J.}\ \bibnamefont {Beenakker}},\ }\bibfield  {title} {\bibinfo {title} {The top-transmon: a hybrid superconducting qubit for parity-protected quantum computation},\ }\href {https://doi.org/10.1088/1367-2630/13/9/095004} {\bibfield  {journal} {\bibinfo  {journal} {New Journal of Physics}\ }\textbf {\bibinfo {volume} {13}},\ \bibinfo {pages} {095004} (\bibinfo {year} {2011})}\BibitemShut {NoStop}%
\bibitem [{\citenamefont {Gambetta}\ \emph {et~al.}(2017)\citenamefont {Gambetta}, \citenamefont {Chow},\ and\ \citenamefont {Steffen}}]{sc_qb_nature_2017}%
  \BibitemOpen
  \bibfield  {author} {\bibinfo {author} {\bibfnamefont {J.~M.}\ \bibnamefont {Gambetta}}, \bibinfo {author} {\bibfnamefont {J.~M.}\ \bibnamefont {Chow}},\ and\ \bibinfo {author} {\bibfnamefont {M.}~\bibnamefont {Steffen}},\ }\bibfield  {title} {\bibinfo {title} {Building logical qubits in a superconducting quantum computing system},\ }\href {https://doi.org/https://doi.org/10.1038/s41534-016-0004-0} {\bibfield  {journal} {\bibinfo  {journal} {npj quantum information}\ }\textbf {\bibinfo {volume} {3}},\ \bibinfo {pages} {2} (\bibinfo {year} {2017})}\BibitemShut {NoStop}%
\bibitem [{\citenamefont {Arute}\ \emph {et~al.}(2019)\citenamefont {Arute}, \citenamefont {Arya}, \citenamefont {Babbush}, \citenamefont {Bacon}, \citenamefont {Bardin}, \citenamefont {Barends}, \citenamefont {Biswas}, \citenamefont {Boixo}, \citenamefont {Brandao}, \citenamefont {Buell}, \citenamefont {Burkett}, \citenamefont {Chen}, \citenamefont {Chen}, \citenamefont {Chiaro}, \citenamefont {Collins}, \citenamefont {Courtney}, \citenamefont {Dunsworth}, \citenamefont {Farhi}, \citenamefont {Foxen}, \citenamefont {Fowler}, \citenamefont {Gidney}, \citenamefont {Giustina}, \citenamefont {Graff}, \citenamefont {Guerin}, \citenamefont {Habegger}, \citenamefont {Harrigan}, \citenamefont {Hartmann}, \citenamefont {Ho}, \citenamefont {Hoffmann}, \citenamefont {Huang}, \citenamefont {Humble}, \citenamefont {Isakov}, \citenamefont {Jeffrey}, \citenamefont {Jiang}, \citenamefont {Kafri}, \citenamefont {Kechedzhi}, \citenamefont {Kelly}, \citenamefont {Klimov}, \citenamefont {Knysh}, \citenamefont {Korotkov},
  \citenamefont {Kostritsa}, \citenamefont {Landhuis}, \citenamefont {Lindmark}, \citenamefont {Lucero}, \citenamefont {Lyakh}, \citenamefont {Mandrà}, \citenamefont {McClean}, \citenamefont {McEwen}, \citenamefont {Megrant}, \citenamefont {Mi}, \citenamefont {Michielsen}, \citenamefont {Mohseni}, \citenamefont {Mutus}, \citenamefont {Naaman}, \citenamefont {Neeley}, \citenamefont {Neill}, \citenamefont {Niu}, \citenamefont {Ostby}, \citenamefont {Petukhov}, \citenamefont {Platt}, \citenamefont {Quintana}, \citenamefont {Rieffel}, \citenamefont {Roushan}, \citenamefont {Rubin}, \citenamefont {Sank}, \citenamefont {Satzinger}, \citenamefont {Smelyanskiy}, \citenamefont {Sung}, \citenamefont {Trevithick}, \citenamefont {Vainsencher}, \citenamefont {Villalonga}, \citenamefont {White}, \citenamefont {Yao}, \citenamefont {Yeh}, \citenamefont {Zalcman}, \citenamefont {Neven},\ and\ \citenamefont {Martinis}}]{google_quantum_sup}%
  \BibitemOpen
  \bibfield  {author} {\bibinfo {author} {\bibfnamefont {F.}~\bibnamefont {Arute}}, \bibinfo {author} {\bibfnamefont {K.}~\bibnamefont {Arya}}, \bibinfo {author} {\bibfnamefont {R.}~\bibnamefont {Babbush}}, \bibinfo {author} {\bibfnamefont {D.}~\bibnamefont {Bacon}}, \bibinfo {author} {\bibfnamefont {J.~C.}\ \bibnamefont {Bardin}}, \bibinfo {author} {\bibfnamefont {R.}~\bibnamefont {Barends}}, \bibinfo {author} {\bibfnamefont {R.}~\bibnamefont {Biswas}}, \bibinfo {author} {\bibfnamefont {S.}~\bibnamefont {Boixo}}, \bibinfo {author} {\bibfnamefont {F.~G.}\ \bibnamefont {Brandao}}, \bibinfo {author} {\bibfnamefont {D.~A.}\ \bibnamefont {Buell}}, \bibinfo {author} {\bibfnamefont {B.}~\bibnamefont {Burkett}}, \bibinfo {author} {\bibfnamefont {Y.}~\bibnamefont {Chen}}, \bibinfo {author} {\bibfnamefont {Z.}~\bibnamefont {Chen}}, \bibinfo {author} {\bibfnamefont {B.}~\bibnamefont {Chiaro}}, \bibinfo {author} {\bibfnamefont {R.}~\bibnamefont {Collins}}, \bibinfo {author} {\bibfnamefont {W.}~\bibnamefont {Courtney}},
  \bibinfo {author} {\bibfnamefont {A.}~\bibnamefont {Dunsworth}}, \bibinfo {author} {\bibfnamefont {E.}~\bibnamefont {Farhi}}, \bibinfo {author} {\bibfnamefont {B.}~\bibnamefont {Foxen}}, \bibinfo {author} {\bibfnamefont {A.}~\bibnamefont {Fowler}}, \bibinfo {author} {\bibfnamefont {C.}~\bibnamefont {Gidney}}, \bibinfo {author} {\bibfnamefont {M.}~\bibnamefont {Giustina}}, \bibinfo {author} {\bibfnamefont {R.}~\bibnamefont {Graff}}, \bibinfo {author} {\bibfnamefont {K.}~\bibnamefont {Guerin}}, \bibinfo {author} {\bibfnamefont {S.}~\bibnamefont {Habegger}}, \bibinfo {author} {\bibfnamefont {M.~P.}\ \bibnamefont {Harrigan}}, \bibinfo {author} {\bibfnamefont {M.~J.}\ \bibnamefont {Hartmann}}, \bibinfo {author} {\bibfnamefont {A.}~\bibnamefont {Ho}}, \bibinfo {author} {\bibfnamefont {M.}~\bibnamefont {Hoffmann}}, \bibinfo {author} {\bibfnamefont {T.}~\bibnamefont {Huang}}, \bibinfo {author} {\bibfnamefont {T.~S.}\ \bibnamefont {Humble}}, \bibinfo {author} {\bibfnamefont {S.~V.}\ \bibnamefont {Isakov}}, \bibinfo
  {author} {\bibfnamefont {E.}~\bibnamefont {Jeffrey}}, \bibinfo {author} {\bibfnamefont {Z.}~\bibnamefont {Jiang}}, \bibinfo {author} {\bibfnamefont {D.}~\bibnamefont {Kafri}}, \bibinfo {author} {\bibfnamefont {K.}~\bibnamefont {Kechedzhi}}, \bibinfo {author} {\bibfnamefont {J.}~\bibnamefont {Kelly}}, \bibinfo {author} {\bibfnamefont {P.~V.}\ \bibnamefont {Klimov}}, \bibinfo {author} {\bibfnamefont {S.}~\bibnamefont {Knysh}}, \bibinfo {author} {\bibfnamefont {A.}~\bibnamefont {Korotkov}}, \bibinfo {author} {\bibfnamefont {F.}~\bibnamefont {Kostritsa}}, \bibinfo {author} {\bibfnamefont {D.}~\bibnamefont {Landhuis}}, \bibinfo {author} {\bibfnamefont {M.}~\bibnamefont {Lindmark}}, \bibinfo {author} {\bibfnamefont {E.}~\bibnamefont {Lucero}}, \bibinfo {author} {\bibfnamefont {D.}~\bibnamefont {Lyakh}}, \bibinfo {author} {\bibfnamefont {S.}~\bibnamefont {Mandrà}}, \bibinfo {author} {\bibfnamefont {J.~R.}\ \bibnamefont {McClean}}, \bibinfo {author} {\bibfnamefont {M.}~\bibnamefont {McEwen}}, \bibinfo {author}
  {\bibfnamefont {A.}~\bibnamefont {Megrant}}, \bibinfo {author} {\bibfnamefont {X.}~\bibnamefont {Mi}}, \bibinfo {author} {\bibfnamefont {K.}~\bibnamefont {Michielsen}}, \bibinfo {author} {\bibfnamefont {M.}~\bibnamefont {Mohseni}}, \bibinfo {author} {\bibfnamefont {J.}~\bibnamefont {Mutus}}, \bibinfo {author} {\bibfnamefont {O.}~\bibnamefont {Naaman}}, \bibinfo {author} {\bibfnamefont {M.}~\bibnamefont {Neeley}}, \bibinfo {author} {\bibfnamefont {C.}~\bibnamefont {Neill}}, \bibinfo {author} {\bibfnamefont {M.~Y.}\ \bibnamefont {Niu}}, \bibinfo {author} {\bibfnamefont {E.}~\bibnamefont {Ostby}}, \bibinfo {author} {\bibfnamefont {A.}~\bibnamefont {Petukhov}}, \bibinfo {author} {\bibfnamefont {J.~C.}\ \bibnamefont {Platt}}, \bibinfo {author} {\bibfnamefont {C.}~\bibnamefont {Quintana}}, \bibinfo {author} {\bibfnamefont {E.~G.}\ \bibnamefont {Rieffel}}, \bibinfo {author} {\bibfnamefont {P.}~\bibnamefont {Roushan}}, \bibinfo {author} {\bibfnamefont {N.~C.}\ \bibnamefont {Rubin}}, \bibinfo {author} {\bibfnamefont
  {D.}~\bibnamefont {Sank}}, \bibinfo {author} {\bibfnamefont {K.~J.}\ \bibnamefont {Satzinger}}, \bibinfo {author} {\bibfnamefont {V.}~\bibnamefont {Smelyanskiy}}, \bibinfo {author} {\bibfnamefont {K.~J.}\ \bibnamefont {Sung}}, \bibinfo {author} {\bibfnamefont {M.~D.}\ \bibnamefont {Trevithick}}, \bibinfo {author} {\bibfnamefont {A.}~\bibnamefont {Vainsencher}}, \bibinfo {author} {\bibfnamefont {B.}~\bibnamefont {Villalonga}}, \bibinfo {author} {\bibfnamefont {T.}~\bibnamefont {White}}, \bibinfo {author} {\bibfnamefont {Z.~J.}\ \bibnamefont {Yao}}, \bibinfo {author} {\bibfnamefont {P.}~\bibnamefont {Yeh}}, \bibinfo {author} {\bibfnamefont {A.}~\bibnamefont {Zalcman}}, \bibinfo {author} {\bibfnamefont {H.}~\bibnamefont {Neven}},\ and\ \bibinfo {author} {\bibfnamefont {J.~M.}\ \bibnamefont {Martinis}},\ }\bibfield  {title} {\bibinfo {title} {Quantum supremacy using a programmable superconducting processor},\ }\href {https://doi.org/https://doi.org/10.1038/s41586-019-1666-5} {\bibfield  {journal} {\bibinfo
  {journal} {Nature}\ }\textbf {\bibinfo {volume} {574}},\ \bibinfo {pages} {505} (\bibinfo {year} {2019})}\BibitemShut {NoStop}%
\bibitem [{\citenamefont {Pogorelov}\ \emph {et~al.}(2021)\citenamefont {Pogorelov}, \citenamefont {Feldker}, \citenamefont {Marciniak}, \citenamefont {Postler}, \citenamefont {Jacob}, \citenamefont {Krieglsteiner}, \citenamefont {Podlesnic}, \citenamefont {Meth}, \citenamefont {Negnevitsky}, \citenamefont {Stadler}, \citenamefont {H\"ofer}, \citenamefont {W\"achter}, \citenamefont {Lakhmanskiy}, \citenamefont {Blatt}, \citenamefont {Schindler},\ and\ \citenamefont {Monz}}]{ion_trap_prx_2021}%
  \BibitemOpen
  \bibfield  {author} {\bibinfo {author} {\bibfnamefont {I.}~\bibnamefont {Pogorelov}}, \bibinfo {author} {\bibfnamefont {T.}~\bibnamefont {Feldker}}, \bibinfo {author} {\bibfnamefont {C.~D.}\ \bibnamefont {Marciniak}}, \bibinfo {author} {\bibfnamefont {L.}~\bibnamefont {Postler}}, \bibinfo {author} {\bibfnamefont {G.}~\bibnamefont {Jacob}}, \bibinfo {author} {\bibfnamefont {O.}~\bibnamefont {Krieglsteiner}}, \bibinfo {author} {\bibfnamefont {V.}~\bibnamefont {Podlesnic}}, \bibinfo {author} {\bibfnamefont {M.}~\bibnamefont {Meth}}, \bibinfo {author} {\bibfnamefont {V.}~\bibnamefont {Negnevitsky}}, \bibinfo {author} {\bibfnamefont {M.}~\bibnamefont {Stadler}}, \bibinfo {author} {\bibfnamefont {B.}~\bibnamefont {H\"ofer}}, \bibinfo {author} {\bibfnamefont {C.}~\bibnamefont {W\"achter}}, \bibinfo {author} {\bibfnamefont {K.}~\bibnamefont {Lakhmanskiy}}, \bibinfo {author} {\bibfnamefont {R.}~\bibnamefont {Blatt}}, \bibinfo {author} {\bibfnamefont {P.}~\bibnamefont {Schindler}},\ and\ \bibinfo {author}
  {\bibfnamefont {T.}~\bibnamefont {Monz}},\ }\bibfield  {title} {\bibinfo {title} {Compact ion-trap quantum computing demonstrator},\ }\href {https://doi.org/10.1103/PRXQuantum.2.020343} {\bibfield  {journal} {\bibinfo  {journal} {PRX Quantum}\ }\textbf {\bibinfo {volume} {2}},\ \bibinfo {pages} {020343} (\bibinfo {year} {2021})}\BibitemShut {NoStop}%
\bibitem [{\citenamefont {Myerson}\ \emph {et~al.}(2008)\citenamefont {Myerson}, \citenamefont {Szwer}, \citenamefont {Webster}, \citenamefont {Allcock}, \citenamefont {Curtis}, \citenamefont {Imreh}, \citenamefont {Sherman}, \citenamefont {Stacey}, \citenamefont {Steane},\ and\ \citenamefont {Lucas}}]{ion_trap_read_out_2008}%
  \BibitemOpen
  \bibfield  {author} {\bibinfo {author} {\bibfnamefont {A.~H.}\ \bibnamefont {Myerson}}, \bibinfo {author} {\bibfnamefont {D.~J.}\ \bibnamefont {Szwer}}, \bibinfo {author} {\bibfnamefont {S.~C.}\ \bibnamefont {Webster}}, \bibinfo {author} {\bibfnamefont {D.~T.~C.}\ \bibnamefont {Allcock}}, \bibinfo {author} {\bibfnamefont {M.~J.}\ \bibnamefont {Curtis}}, \bibinfo {author} {\bibfnamefont {G.}~\bibnamefont {Imreh}}, \bibinfo {author} {\bibfnamefont {J.~A.}\ \bibnamefont {Sherman}}, \bibinfo {author} {\bibfnamefont {D.~N.}\ \bibnamefont {Stacey}}, \bibinfo {author} {\bibfnamefont {A.~M.}\ \bibnamefont {Steane}},\ and\ \bibinfo {author} {\bibfnamefont {D.~M.}\ \bibnamefont {Lucas}},\ }\bibfield  {title} {\bibinfo {title} {High-fidelity readout of trapped-ion qubits},\ }\href {https://doi.org/10.1103/PhysRevLett.100.200502} {\bibfield  {journal} {\bibinfo  {journal} {Phys. Rev. Lett.}\ }\textbf {\bibinfo {volume} {100}},\ \bibinfo {pages} {200502} (\bibinfo {year} {2008})}\BibitemShut {NoStop}%
\bibitem [{\citenamefont {Wright}\ \emph {et~al.}(2019)\citenamefont {Wright}, \citenamefont {Beck}, \citenamefont {Debnath}, \citenamefont {Amini}, \citenamefont {Nam}, \citenamefont {Grzesiak}, \citenamefont {Chen}, \citenamefont {Pisenti}, \citenamefont {Chmielewski}, \citenamefont {Collins}, \citenamefont {Hudek}, \citenamefont {Mizrahi}, \citenamefont {Wong-Campos}, \citenamefont {Allen}, \citenamefont {Apisdorf}, \citenamefont {Solomon}, \citenamefont {Williams}, \citenamefont {Ducore}, \citenamefont {Blinov}, \citenamefont {Kreikemeier}, \citenamefont {Chaplin}, \citenamefont {Keesan}, \citenamefont {Monroe},\ and\ \citenamefont {Kim}}]{ion_trap_wright2019}%
  \BibitemOpen
  \bibfield  {author} {\bibinfo {author} {\bibfnamefont {K.}~\bibnamefont {Wright}}, \bibinfo {author} {\bibfnamefont {K.~M.}\ \bibnamefont {Beck}}, \bibinfo {author} {\bibfnamefont {S.}~\bibnamefont {Debnath}}, \bibinfo {author} {\bibfnamefont {J.}~\bibnamefont {Amini}}, \bibinfo {author} {\bibfnamefont {Y.}~\bibnamefont {Nam}}, \bibinfo {author} {\bibfnamefont {N.}~\bibnamefont {Grzesiak}}, \bibinfo {author} {\bibfnamefont {J.-S.}\ \bibnamefont {Chen}}, \bibinfo {author} {\bibfnamefont {N.}~\bibnamefont {Pisenti}}, \bibinfo {author} {\bibfnamefont {M.}~\bibnamefont {Chmielewski}}, \bibinfo {author} {\bibfnamefont {C.}~\bibnamefont {Collins}}, \bibinfo {author} {\bibfnamefont {K.~M.}\ \bibnamefont {Hudek}}, \bibinfo {author} {\bibfnamefont {J.}~\bibnamefont {Mizrahi}}, \bibinfo {author} {\bibfnamefont {J.~D.}\ \bibnamefont {Wong-Campos}}, \bibinfo {author} {\bibfnamefont {S.}~\bibnamefont {Allen}}, \bibinfo {author} {\bibfnamefont {J.}~\bibnamefont {Apisdorf}}, \bibinfo {author} {\bibfnamefont
  {P.}~\bibnamefont {Solomon}}, \bibinfo {author} {\bibfnamefont {M.}~\bibnamefont {Williams}}, \bibinfo {author} {\bibfnamefont {A.~M.}\ \bibnamefont {Ducore}}, \bibinfo {author} {\bibfnamefont {A.}~\bibnamefont {Blinov}}, \bibinfo {author} {\bibfnamefont {S.~M.}\ \bibnamefont {Kreikemeier}}, \bibinfo {author} {\bibfnamefont {V.}~\bibnamefont {Chaplin}}, \bibinfo {author} {\bibfnamefont {M.}~\bibnamefont {Keesan}}, \bibinfo {author} {\bibfnamefont {C.}~\bibnamefont {Monroe}},\ and\ \bibinfo {author} {\bibfnamefont {J.}~\bibnamefont {Kim}},\ }\bibfield  {title} {\bibinfo {title} {Benchmarking an 11-qubit quantum computer},\ }\href {https://doi.org/https://doi.org/10.1038/s41467-019-13534-2} {\bibfield  {journal} {\bibinfo  {journal} {Nature communications}\ }\textbf {\bibinfo {volume} {10}},\ \bibinfo {pages} {5464} (\bibinfo {year} {2019})}\BibitemShut {NoStop}%
\bibitem [{\citenamefont {Knill}\ \emph {et~al.}(2001)\citenamefont {Knill}, \citenamefont {Laflamme},\ and\ \citenamefont {Milburn}}]{loqc_2001}%
  \BibitemOpen
  \bibfield  {author} {\bibinfo {author} {\bibfnamefont {E.}~\bibnamefont {Knill}}, \bibinfo {author} {\bibfnamefont {R.}~\bibnamefont {Laflamme}},\ and\ \bibinfo {author} {\bibfnamefont {G.~J.}\ \bibnamefont {Milburn}},\ }\bibfield  {title} {\bibinfo {title} {A scheme for efficient quantum computation with linear optics},\ }\href {https://doi.org/https://doi.org/10.1038/35051009} {\bibfield  {journal} {\bibinfo  {journal} {nature}\ }\textbf {\bibinfo {volume} {409}},\ \bibinfo {pages} {46} (\bibinfo {year} {2001})}\BibitemShut {NoStop}%
\bibitem [{\citenamefont {Kok}\ \emph {et~al.}(2007)\citenamefont {Kok}, \citenamefont {Munro}, \citenamefont {Nemoto}, \citenamefont {Ralph}, \citenamefont {Dowling},\ and\ \citenamefont {Milburn}}]{loqc_ph_qb_2007}%
  \BibitemOpen
  \bibfield  {author} {\bibinfo {author} {\bibfnamefont {P.}~\bibnamefont {Kok}}, \bibinfo {author} {\bibfnamefont {W.~J.}\ \bibnamefont {Munro}}, \bibinfo {author} {\bibfnamefont {K.}~\bibnamefont {Nemoto}}, \bibinfo {author} {\bibfnamefont {T.~C.}\ \bibnamefont {Ralph}}, \bibinfo {author} {\bibfnamefont {J.~P.}\ \bibnamefont {Dowling}},\ and\ \bibinfo {author} {\bibfnamefont {G.~J.}\ \bibnamefont {Milburn}},\ }\bibfield  {title} {\bibinfo {title} {Linear optical quantum computing with photonic qubits},\ }\href {https://doi.org/10.1103/RevModPhys.79.135} {\bibfield  {journal} {\bibinfo  {journal} {Rev. Mod. Phys.}\ }\textbf {\bibinfo {volume} {79}},\ \bibinfo {pages} {135} (\bibinfo {year} {2007})}\BibitemShut {NoStop}%
\bibitem [{\citenamefont {Madsen}\ \emph {et~al.}(2022)\citenamefont {Madsen}, \citenamefont {Laudenbach}, \citenamefont {Askarani}, \citenamefont {Rortais}, \citenamefont {Vincent}, \citenamefont {Bulmer}, \citenamefont {Miatto}, \citenamefont {Neuhaus}, \citenamefont {Helt}, \citenamefont {Collins}, \citenamefont {Lita}, \citenamefont {Gerrits}, \citenamefont {Nam}, \citenamefont {Vaidya}, \citenamefont {Menotti}, \citenamefont {Dhand}, \citenamefont {Vernon}, \citenamefont {Quesada},\ and\ \citenamefont {Lavoie}}]{xanadu_quantum_advantage}%
  \BibitemOpen
  \bibfield  {author} {\bibinfo {author} {\bibfnamefont {L.~S.}\ \bibnamefont {Madsen}}, \bibinfo {author} {\bibfnamefont {F.}~\bibnamefont {Laudenbach}}, \bibinfo {author} {\bibfnamefont {M.~F.}\ \bibnamefont {Askarani}}, \bibinfo {author} {\bibfnamefont {F.}~\bibnamefont {Rortais}}, \bibinfo {author} {\bibfnamefont {T.}~\bibnamefont {Vincent}}, \bibinfo {author} {\bibfnamefont {J.~F.}\ \bibnamefont {Bulmer}}, \bibinfo {author} {\bibfnamefont {F.~M.}\ \bibnamefont {Miatto}}, \bibinfo {author} {\bibfnamefont {L.}~\bibnamefont {Neuhaus}}, \bibinfo {author} {\bibfnamefont {L.~G.}\ \bibnamefont {Helt}}, \bibinfo {author} {\bibfnamefont {M.~J.}\ \bibnamefont {Collins}}, \bibinfo {author} {\bibfnamefont {A.~E.}\ \bibnamefont {Lita}}, \bibinfo {author} {\bibfnamefont {T.}~\bibnamefont {Gerrits}}, \bibinfo {author} {\bibfnamefont {S.~W.}\ \bibnamefont {Nam}}, \bibinfo {author} {\bibfnamefont {V.~D.}\ \bibnamefont {Vaidya}}, \bibinfo {author} {\bibfnamefont {M.}~\bibnamefont {Menotti}}, \bibinfo {author}
  {\bibfnamefont {I.}~\bibnamefont {Dhand}}, \bibinfo {author} {\bibfnamefont {Z.}~\bibnamefont {Vernon}}, \bibinfo {author} {\bibfnamefont {N.}~\bibnamefont {Quesada}},\ and\ \bibinfo {author} {\bibfnamefont {J.}~\bibnamefont {Lavoie}},\ }\bibfield  {title} {\bibinfo {title} {Quantum computational advantage with a programmable photonic processor},\ }\href {https://doi.org/https://doi.org/10.1038/s41586-022-04725-x} {\bibfield  {journal} {\bibinfo  {journal} {Nature}\ }\textbf {\bibinfo {volume} {606}},\ \bibinfo {pages} {75} (\bibinfo {year} {2022})}\BibitemShut {NoStop}%
\bibitem [{\citenamefont {Northup}\ and\ \citenamefont {Blatt}(2014{\natexlab{b}})}]{northup2014quantum}%
  \BibitemOpen
  \bibfield  {author} {\bibinfo {author} {\bibfnamefont {T.}~\bibnamefont {Northup}}\ and\ \bibinfo {author} {\bibfnamefont {R.}~\bibnamefont {Blatt}},\ }\bibfield  {title} {\bibinfo {title} {Quantum information transfer using photons},\ }\href {https://doi.org/10.1146/annurev-conmatphys-030212-184248} {\bibfield  {journal} {\bibinfo  {journal} {Nature photonics}\ }\textbf {\bibinfo {volume} {8}},\ \bibinfo {pages} {356} (\bibinfo {year} {2014}{\natexlab{b}})}\BibitemShut {NoStop}%
\bibitem [{\citenamefont {Brum}\ and\ \citenamefont {Hawrylak}(1997)}]{coupled_qd_qb_1997}%
  \BibitemOpen
  \bibfield  {author} {\bibinfo {author} {\bibfnamefont {J.~A.}\ \bibnamefont {Brum}}\ and\ \bibinfo {author} {\bibfnamefont {P.}~\bibnamefont {Hawrylak}},\ }\bibfield  {title} {\bibinfo {title} {Coupled quantum dots as quantum exclusive-{OR} gate},\ }\href {https://doi.org/https://doi.org/10.1006/spmi.1996.0263} {\bibfield  {journal} {\bibinfo  {journal} {Superlattices and Microstructures}\ }\textbf {\bibinfo {volume} {22}},\ \bibinfo {pages} {431} (\bibinfo {year} {1997})}\BibitemShut {NoStop}%
\bibitem [{\citenamefont {Loss}\ and\ \citenamefont {DiVincenzo}(1998)}]{qc_qd_1997}%
  \BibitemOpen
  \bibfield  {author} {\bibinfo {author} {\bibfnamefont {D.}~\bibnamefont {Loss}}\ and\ \bibinfo {author} {\bibfnamefont {D.~P.}\ \bibnamefont {DiVincenzo}},\ }\bibfield  {title} {\bibinfo {title} {Quantum computation with quantum dots},\ }\href {https://doi.org/10.1103/PhysRevA.57.120} {\bibfield  {journal} {\bibinfo  {journal} {Phys. Rev. A}\ }\textbf {\bibinfo {volume} {57}},\ \bibinfo {pages} {120} (\bibinfo {year} {1998})}\BibitemShut {NoStop}%
\bibitem [{\citenamefont {Alfieri}\ \emph {et~al.}(2023)\citenamefont {Alfieri}, \citenamefont {Anantharaman}, \citenamefont {Zhang},\ and\ \citenamefont {Jariwala}}]{nano_qi_2023}%
  \BibitemOpen
  \bibfield  {author} {\bibinfo {author} {\bibfnamefont {A.}~\bibnamefont {Alfieri}}, \bibinfo {author} {\bibfnamefont {S.~B.}\ \bibnamefont {Anantharaman}}, \bibinfo {author} {\bibfnamefont {H.}~\bibnamefont {Zhang}},\ and\ \bibinfo {author} {\bibfnamefont {D.}~\bibnamefont {Jariwala}},\ }\bibfield  {title} {\bibinfo {title} {Nanomaterials for quantum information science and engineering},\ }\href {https://doi.org/https://doi.org/10.1002/adma.202109621} {\bibfield  {journal} {\bibinfo  {journal} {Advanced Materials}\ }\textbf {\bibinfo {volume} {35}},\ \bibinfo {pages} {2109621} (\bibinfo {year} {2023})}\BibitemShut {NoStop}%
\bibitem [{\citenamefont {Kloeffel}\ and\ \citenamefont {Loss}(2013{\natexlab{b}})}]{spin_qd_qubit_2013}%
  \BibitemOpen
  \bibfield  {author} {\bibinfo {author} {\bibfnamefont {C.}~\bibnamefont {Kloeffel}}\ and\ \bibinfo {author} {\bibfnamefont {D.}~\bibnamefont {Loss}},\ }\bibfield  {title} {\bibinfo {title} {Prospects for spin-based quantum computing in quantum dots},\ }\href {https://doi.org/10.1146/annurev-conmatphys-030212-184248} {\bibfield  {journal} {\bibinfo  {journal} {Annual Review of Condensed Matter Physics}\ }\textbf {\bibinfo {volume} {4}},\ \bibinfo {pages} {51} (\bibinfo {year} {2013}{\natexlab{b}})}\BibitemShut {NoStop}%
\bibitem [{\citenamefont {Kobayashi}\ \emph {et~al.}(2021)\citenamefont {Kobayashi}, \citenamefont {Salfi}, \citenamefont {Chua}, \citenamefont {Van Der~Heijden}, \citenamefont {House}, \citenamefont {Culcer}, \citenamefont {Hutchison}, \citenamefont {Johnson}, \citenamefont {McCallum}, \citenamefont {Riemann}, \citenamefont {Abrosimov}, \citenamefont {Becker}, \citenamefont {Pohl}, \citenamefont {Simmons},\ and\ \citenamefont {Rogge}}]{so_qubit_si_2021}%
  \BibitemOpen
  \bibfield  {author} {\bibinfo {author} {\bibfnamefont {T.}~\bibnamefont {Kobayashi}}, \bibinfo {author} {\bibfnamefont {J.}~\bibnamefont {Salfi}}, \bibinfo {author} {\bibfnamefont {C.}~\bibnamefont {Chua}}, \bibinfo {author} {\bibfnamefont {J.}~\bibnamefont {Van Der~Heijden}}, \bibinfo {author} {\bibfnamefont {M.~G.}\ \bibnamefont {House}}, \bibinfo {author} {\bibfnamefont {D.}~\bibnamefont {Culcer}}, \bibinfo {author} {\bibfnamefont {W.~D.}\ \bibnamefont {Hutchison}}, \bibinfo {author} {\bibfnamefont {B.~C.}\ \bibnamefont {Johnson}}, \bibinfo {author} {\bibfnamefont {J.~C.}\ \bibnamefont {McCallum}}, \bibinfo {author} {\bibfnamefont {H.}~\bibnamefont {Riemann}}, \bibinfo {author} {\bibfnamefont {N.~V.}\ \bibnamefont {Abrosimov}}, \bibinfo {author} {\bibfnamefont {P.}~\bibnamefont {Becker}}, \bibinfo {author} {\bibfnamefont {H.-J.}\ \bibnamefont {Pohl}}, \bibinfo {author} {\bibfnamefont {M.~Y.}\ \bibnamefont {Simmons}},\ and\ \bibinfo {author} {\bibfnamefont {S.}~\bibnamefont {Rogge}},\ }\bibfield  {title}
  {\bibinfo {title} {Engineering long spin coherence times of spin--orbit qubits in silicon},\ }\href {https://doi.org/https://doi.org/10.1038/s41563-020-0743-3} {\bibfield  {journal} {\bibinfo  {journal} {Nature Materials}\ }\textbf {\bibinfo {volume} {20}},\ \bibinfo {pages} {38} (\bibinfo {year} {2021})}\BibitemShut {NoStop}%
\bibitem [{\citenamefont {Koppens}\ \emph {et~al.}(2006)\citenamefont {Koppens}, \citenamefont {Buizert}, \citenamefont {Tielrooij}, \citenamefont {Vink}, \citenamefont {Nowack}, \citenamefont {Meunier}, \citenamefont {Kouwenhoven},\ and\ \citenamefont {Vandersypen}}]{coherent_spin_qd_2006}%
  \BibitemOpen
  \bibfield  {author} {\bibinfo {author} {\bibfnamefont {F.~H.}\ \bibnamefont {Koppens}}, \bibinfo {author} {\bibfnamefont {C.}~\bibnamefont {Buizert}}, \bibinfo {author} {\bibfnamefont {K.-J.}\ \bibnamefont {Tielrooij}}, \bibinfo {author} {\bibfnamefont {I.~T.}\ \bibnamefont {Vink}}, \bibinfo {author} {\bibfnamefont {K.~C.}\ \bibnamefont {Nowack}}, \bibinfo {author} {\bibfnamefont {T.}~\bibnamefont {Meunier}}, \bibinfo {author} {\bibfnamefont {L.}~\bibnamefont {Kouwenhoven}},\ and\ \bibinfo {author} {\bibfnamefont {L.}~\bibnamefont {Vandersypen}},\ }\bibfield  {title} {\bibinfo {title} {Driven coherent oscillations of a single electron spin in a quantum dot},\ }\href {https://doi.org/https://doi.org/10.1038/nature05065} {\bibfield  {journal} {\bibinfo  {journal} {Nature}\ }\textbf {\bibinfo {volume} {442}},\ \bibinfo {pages} {766} (\bibinfo {year} {2006})}\BibitemShut {NoStop}%
\bibitem [{\citenamefont {Muhonen}\ \emph {et~al.}(2014)\citenamefont {Muhonen}, \citenamefont {Dehollain}, \citenamefont {Laucht}, \citenamefont {Hudson}, \citenamefont {Kalra}, \citenamefont {Sekiguchi}, \citenamefont {Itoh}, \citenamefont {Jamieson}, \citenamefont {McCallum}, \citenamefont {Dzurak},\ and\ \citenamefont {Morello}}]{store_spin_si_2014}%
  \BibitemOpen
  \bibfield  {author} {\bibinfo {author} {\bibfnamefont {J.~T.}\ \bibnamefont {Muhonen}}, \bibinfo {author} {\bibfnamefont {J.~P.}\ \bibnamefont {Dehollain}}, \bibinfo {author} {\bibfnamefont {A.}~\bibnamefont {Laucht}}, \bibinfo {author} {\bibfnamefont {F.~E.}\ \bibnamefont {Hudson}}, \bibinfo {author} {\bibfnamefont {R.}~\bibnamefont {Kalra}}, \bibinfo {author} {\bibfnamefont {T.}~\bibnamefont {Sekiguchi}}, \bibinfo {author} {\bibfnamefont {K.~M.}\ \bibnamefont {Itoh}}, \bibinfo {author} {\bibfnamefont {D.~N.}\ \bibnamefont {Jamieson}}, \bibinfo {author} {\bibfnamefont {J.~C.}\ \bibnamefont {McCallum}}, \bibinfo {author} {\bibfnamefont {A.~S.}\ \bibnamefont {Dzurak}},\ and\ \bibinfo {author} {\bibfnamefont {A.}~\bibnamefont {Morello}},\ }\bibfield  {title} {\bibinfo {title} {Storing quantum information for 30 seconds in a nanoelectronic device},\ }\href {https://doi.org/https://doi.org/10.1038/nnano.2014.211} {\bibfield  {journal} {\bibinfo  {journal} {Nature nanotechnology}\ }\textbf {\bibinfo {volume}
  {9}},\ \bibinfo {pages} {986} (\bibinfo {year} {2014})}\BibitemShut {NoStop}%
\bibitem [{\citenamefont {Petta}\ \emph {et~al.}(2005)\citenamefont {Petta}, \citenamefont {Johnson}, \citenamefont {Taylor}, \citenamefont {Laird}, \citenamefont {Yacoby}, \citenamefont {Lukin}, \citenamefont {Marcus}, \citenamefont {Hanson},\ and\ \citenamefont {Gossard}}]{Petta_STQB_2005}%
  \BibitemOpen
  \bibfield  {author} {\bibinfo {author} {\bibfnamefont {J.~R.}\ \bibnamefont {Petta}}, \bibinfo {author} {\bibfnamefont {A.~C.}\ \bibnamefont {Johnson}}, \bibinfo {author} {\bibfnamefont {J.~M.}\ \bibnamefont {Taylor}}, \bibinfo {author} {\bibfnamefont {E.~A.}\ \bibnamefont {Laird}}, \bibinfo {author} {\bibfnamefont {A.}~\bibnamefont {Yacoby}}, \bibinfo {author} {\bibfnamefont {M.~D.}\ \bibnamefont {Lukin}}, \bibinfo {author} {\bibfnamefont {C.~M.}\ \bibnamefont {Marcus}}, \bibinfo {author} {\bibfnamefont {M.~P.}\ \bibnamefont {Hanson}},\ and\ \bibinfo {author} {\bibfnamefont {A.~C.}\ \bibnamefont {Gossard}},\ }\bibfield  {title} {\bibinfo {title} {Coherent manipulation of coupled electron spins in semiconductor quantum dots},\ }\href {https://doi.org/10.1126/science.1116955} {\bibfield  {journal} {\bibinfo  {journal} {Science}\ }\textbf {\bibinfo {volume} {309}},\ \bibinfo {pages} {2180} (\bibinfo {year} {2005})}\BibitemShut {NoStop}%
\bibitem [{\citenamefont {Ciorga}\ \emph {et~al.}(2000)\citenamefont {Ciorga}, \citenamefont {Sachrajda}, \citenamefont {Hawrylak}, \citenamefont {Gould}, \citenamefont {Zawadzki}, \citenamefont {Jullian}, \citenamefont {Feng},\ and\ \citenamefont {Wasilewski}}]{Ciorga2000}%
  \BibitemOpen
  \bibfield  {author} {\bibinfo {author} {\bibfnamefont {M.}~\bibnamefont {Ciorga}}, \bibinfo {author} {\bibfnamefont {A.~S.}\ \bibnamefont {Sachrajda}}, \bibinfo {author} {\bibfnamefont {P.}~\bibnamefont {Hawrylak}}, \bibinfo {author} {\bibfnamefont {C.}~\bibnamefont {Gould}}, \bibinfo {author} {\bibfnamefont {P.}~\bibnamefont {Zawadzki}}, \bibinfo {author} {\bibfnamefont {S.}~\bibnamefont {Jullian}}, \bibinfo {author} {\bibfnamefont {Y.}~\bibnamefont {Feng}},\ and\ \bibinfo {author} {\bibfnamefont {Z.}~\bibnamefont {Wasilewski}},\ }\bibfield  {title} {\bibinfo {title} {Addition spectrum of a lateral dot from coulomb and spin-blockade spectroscopy},\ }\href {https://doi.org/10.1103/PhysRevB.61.R16315} {\bibfield  {journal} {\bibinfo  {journal} {Phys. Rev. B}\ }\textbf {\bibinfo {volume} {61}},\ \bibinfo {pages} {R16315} (\bibinfo {year} {2000})}\BibitemShut {NoStop}%
\bibitem [{\citenamefont {Gaudreau}\ \emph {et~al.}(2006)\citenamefont {Gaudreau}, \citenamefont {Studenikin}, \citenamefont {Sachrajda}, \citenamefont {Zawadzki}, \citenamefont {Kam}, \citenamefont {Lapointe}, \citenamefont {Korkusinski},\ and\ \citenamefont {Hawrylak}}]{GaudreauPRL2006}%
  \BibitemOpen
  \bibfield  {author} {\bibinfo {author} {\bibfnamefont {L.}~\bibnamefont {Gaudreau}}, \bibinfo {author} {\bibfnamefont {S.~A.}\ \bibnamefont {Studenikin}}, \bibinfo {author} {\bibfnamefont {A.~S.}\ \bibnamefont {Sachrajda}}, \bibinfo {author} {\bibfnamefont {P.}~\bibnamefont {Zawadzki}}, \bibinfo {author} {\bibfnamefont {A.}~\bibnamefont {Kam}}, \bibinfo {author} {\bibfnamefont {J.}~\bibnamefont {Lapointe}}, \bibinfo {author} {\bibfnamefont {M.}~\bibnamefont {Korkusinski}},\ and\ \bibinfo {author} {\bibfnamefont {P.}~\bibnamefont {Hawrylak}},\ }\bibfield  {title} {\bibinfo {title} {Stability diagram of a few-electron triple dot},\ }\href {https://doi.org/10.1103/PhysRevLett.97.036807} {\bibfield  {journal} {\bibinfo  {journal} {Phys. Rev. Lett.}\ }\textbf {\bibinfo {volume} {97}},\ \bibinfo {pages} {036807} (\bibinfo {year} {2006})}\BibitemShut {NoStop}%
\bibitem [{\citenamefont {Pioro-Ladri\`ere}\ \emph {et~al.}(2003)\citenamefont {Pioro-Ladri\`ere}, \citenamefont {Ciorga}, \citenamefont {Lapointe}, \citenamefont {Zawadzki}, \citenamefont {Korkusi\ifmmode~\acute{n}\else \'{n}\fi{}ski}, \citenamefont {Hawrylak},\ and\ \citenamefont {Sachrajda}}]{PioroPRL2021}%
  \BibitemOpen
  \bibfield  {author} {\bibinfo {author} {\bibfnamefont {M.}~\bibnamefont {Pioro-Ladri\`ere}}, \bibinfo {author} {\bibfnamefont {M.}~\bibnamefont {Ciorga}}, \bibinfo {author} {\bibfnamefont {J.}~\bibnamefont {Lapointe}}, \bibinfo {author} {\bibfnamefont {P.}~\bibnamefont {Zawadzki}}, \bibinfo {author} {\bibfnamefont {M.}~\bibnamefont {Korkusi\ifmmode~\acute{n}\else \'{n}\fi{}ski}}, \bibinfo {author} {\bibfnamefont {P.}~\bibnamefont {Hawrylak}},\ and\ \bibinfo {author} {\bibfnamefont {A.~S.}\ \bibnamefont {Sachrajda}},\ }\bibfield  {title} {\bibinfo {title} {Spin-blockade spectroscopy of a two-level artificial molecule},\ }\href {https://doi.org/10.1103/PhysRevLett.91.026803} {\bibfield  {journal} {\bibinfo  {journal} {Phys. Rev. Lett.}\ }\textbf {\bibinfo {volume} {91}},\ \bibinfo {pages} {026803} (\bibinfo {year} {2003})}\BibitemShut {NoStop}%
\bibitem [{\citenamefont {Haldane}(1983)}]{haldane_physlett_1983}%
  \BibitemOpen
  \bibfield  {author} {\bibinfo {author} {\bibfnamefont {F.}~\bibnamefont {Haldane}},\ }\bibfield  {title} {\bibinfo {title} {Continuum dynamics of the 1-d heisenberg antiferromagnet: Identification with the o(3) nonlinear sigma model},\ }\href {https://doi.org/https://doi.org/10.1016/0375-9601(83)90631-X} {\bibfield  {journal} {\bibinfo  {journal} {Physics Letters A}\ }\textbf {\bibinfo {volume} {93}},\ \bibinfo {pages} {464} (\bibinfo {year} {1983})}\BibitemShut {NoStop}%
\bibitem [{\citenamefont {Haldane}(2017)}]{haldane_rev_mod_phys_2017}%
  \BibitemOpen
  \bibfield  {author} {\bibinfo {author} {\bibfnamefont {F.~D.~M.}\ \bibnamefont {Haldane}},\ }\bibfield  {title} {\bibinfo {title} {Nobel lecture: Topological quantum matter},\ }\href {https://doi.org/10.1103/RevModPhys.89.040502} {\bibfield  {journal} {\bibinfo  {journal} {Rev. Mod. Phys.}\ }\textbf {\bibinfo {volume} {89}},\ \bibinfo {pages} {040502} (\bibinfo {year} {2017})}\BibitemShut {NoStop}%
\bibitem [{\citenamefont {Affleck}\ \emph {et~al.}(1987)\citenamefont {Affleck}, \citenamefont {Kennedy}, \citenamefont {Lieb},\ and\ \citenamefont {Tasaki}}]{affleck_prl_1987}%
  \BibitemOpen
  \bibfield  {author} {\bibinfo {author} {\bibfnamefont {I.}~\bibnamefont {Affleck}}, \bibinfo {author} {\bibfnamefont {T.}~\bibnamefont {Kennedy}}, \bibinfo {author} {\bibfnamefont {E.~H.}\ \bibnamefont {Lieb}},\ and\ \bibinfo {author} {\bibfnamefont {H.}~\bibnamefont {Tasaki}},\ }\bibfield  {title} {\bibinfo {title} {Rigorous results on valence-bond ground states in antiferromagnets},\ }\href {https://doi.org/10.1103/PhysRevLett.59.799} {\bibfield  {journal} {\bibinfo  {journal} {Phys. Rev. Lett.}\ }\textbf {\bibinfo {volume} {59}},\ \bibinfo {pages} {799} (\bibinfo {year} {1987})}\BibitemShut {NoStop}%
\bibitem [{\citenamefont {Jaworowski}\ \emph {et~al.}(2017)\citenamefont {Jaworowski}, \citenamefont {Rogers}, \citenamefont {Grabowski},\ and\ \citenamefont {Hawrylak}}]{jaworowski_scirep_2017}%
  \BibitemOpen
  \bibfield  {author} {\bibinfo {author} {\bibfnamefont {B.}~\bibnamefont {Jaworowski}}, \bibinfo {author} {\bibfnamefont {N.}~\bibnamefont {Rogers}}, \bibinfo {author} {\bibfnamefont {M.}~\bibnamefont {Grabowski}},\ and\ \bibinfo {author} {\bibfnamefont {P.}~\bibnamefont {Hawrylak}},\ }\bibfield  {title} {\bibinfo {title} {Macroscopic singlet-triplet qubit in synthetic spin-one chain in semiconductor nanowires},\ }\href {https://doi.org/10.1038/s41598-017-05655-9} {\bibfield  {journal} {\bibinfo  {journal} {Scientific Reports}\ }\textbf {\bibinfo {volume} {7}},\ \bibinfo {pages} {5529} (\bibinfo {year} {2017})}\BibitemShut {NoStop}%
\bibitem [{\citenamefont {Shim}\ \emph {et~al.}(2010)\citenamefont {Shim}, \citenamefont {Sharma}, \citenamefont {Hsieh},\ and\ \citenamefont {Hawrylak}}]{shim-hawrylak-ssc2010}%
  \BibitemOpen
  \bibfield  {author} {\bibinfo {author} {\bibfnamefont {Y.-P.}\ \bibnamefont {Shim}}, \bibinfo {author} {\bibfnamefont {A.}~\bibnamefont {Sharma}}, \bibinfo {author} {\bibfnamefont {C.-Y.}\ \bibnamefont {Hsieh}},\ and\ \bibinfo {author} {\bibfnamefont {P.}~\bibnamefont {Hawrylak}},\ }\bibfield  {title} {\bibinfo {title} {Artificial haldane gap material on a semiconductor chip},\ }\href@noop {} {\bibfield  {journal} {\bibinfo  {journal} {Solid State Comm}\ }\textbf {\bibinfo {volume} {150}},\ \bibinfo {pages} {2065} (\bibinfo {year} {2010})}\BibitemShut {NoStop}%
\bibitem [{\citenamefont {Laferrière}\ \emph {et~al.}(2021)\citenamefont {Laferrière}, \citenamefont {Yeung}, \citenamefont {Korkusinski}, \citenamefont {Poole}, \citenamefont {Williams}, \citenamefont {Dalacu}, \citenamefont {Manalo}, \citenamefont {Cygorek}, \citenamefont {Altintas},\ and\ \citenamefont {Hawrylak}}]{lafferiere_apl_2021}%
  \BibitemOpen
  \bibfield  {author} {\bibinfo {author} {\bibfnamefont {P.}~\bibnamefont {Laferrière}}, \bibinfo {author} {\bibfnamefont {E.}~\bibnamefont {Yeung}}, \bibinfo {author} {\bibfnamefont {M.}~\bibnamefont {Korkusinski}}, \bibinfo {author} {\bibfnamefont {P.~J.}\ \bibnamefont {Poole}}, \bibinfo {author} {\bibfnamefont {R.~L.}\ \bibnamefont {Williams}}, \bibinfo {author} {\bibfnamefont {D.}~\bibnamefont {Dalacu}}, \bibinfo {author} {\bibfnamefont {J.}~\bibnamefont {Manalo}}, \bibinfo {author} {\bibfnamefont {M.}~\bibnamefont {Cygorek}}, \bibinfo {author} {\bibfnamefont {A.}~\bibnamefont {Altintas}},\ and\ \bibinfo {author} {\bibfnamefont {P.}~\bibnamefont {Hawrylak}},\ }\bibfield  {title} {\bibinfo {title} {Systematic study of the emission spectra of nanowire quantum dots},\ }\href {https://doi.org/10.1063/5.0045880} {\bibfield  {journal} {\bibinfo  {journal} {Applied Physics Letters}\ }\textbf {\bibinfo {volume} {118}},\ \bibinfo {pages} {161107} (\bibinfo {year} {2021})}\BibitemShut {NoStop}%
\bibitem [{\citenamefont {Manalo}\ \emph {et~al.}(2021)\citenamefont {Manalo}, \citenamefont {Cygorek}, \citenamefont {Altintas},\ and\ \citenamefont {Hawrylak}}]{manalo_prb_2021}%
  \BibitemOpen
  \bibfield  {author} {\bibinfo {author} {\bibfnamefont {J.}~\bibnamefont {Manalo}}, \bibinfo {author} {\bibfnamefont {M.}~\bibnamefont {Cygorek}}, \bibinfo {author} {\bibfnamefont {A.}~\bibnamefont {Altintas}},\ and\ \bibinfo {author} {\bibfnamefont {P.}~\bibnamefont {Hawrylak}},\ }\bibfield  {title} {\bibinfo {title} {Electronic and magnetic properties of many-electron complexes in charged {In}{$\text{As}_{x}$}{P}$_{1-x}$ quantum dots in {InP} nanowires},\ }\href {https://doi.org/10.1103/PhysRevB.104.125402} {\bibfield  {journal} {\bibinfo  {journal} {Phys. Rev. B}\ }\textbf {\bibinfo {volume} {104}},\ \bibinfo {pages} {125402} (\bibinfo {year} {2021})}\BibitemShut {NoStop}%
\bibitem [{\citenamefont {Allami}\ \emph {et~al.}(2025)\citenamefont {Allami}, \citenamefont {Miravet}, \citenamefont {Korkusinski},\ and\ \citenamefont {Hawrylak}}]{AllamiPRBTwoQubits}%
  \BibitemOpen
  \bibfield  {author} {\bibinfo {author} {\bibfnamefont {H.}~\bibnamefont {Allami}}, \bibinfo {author} {\bibfnamefont {D.}~\bibnamefont {Miravet}}, \bibinfo {author} {\bibfnamefont {M.}~\bibnamefont {Korkusinski}},\ and\ \bibinfo {author} {\bibfnamefont {P.}~\bibnamefont {Hawrylak}},\ }\bibfield  {title} {\bibinfo {title} {Two-qubit gate with macroscopic singlet-triplet qubits in synthetic spin-one chains in {InAsP} quantum dot nanowires},\ }\href {https://doi.org/10.1103/PhysRevB.111.115403} {\bibfield  {journal} {\bibinfo  {journal} {Phys. Rev. B}\ }\textbf {\bibinfo {volume} {111}},\ \bibinfo {pages} {115403} (\bibinfo {year} {2025})}\BibitemShut {NoStop}%
\bibitem [{\citenamefont {Guclu}\ \emph {et~al.}(2014)\citenamefont {Guclu}, \citenamefont {Potasz}, \citenamefont {Korkusinski},\ and\ \citenamefont {Hawrylak}}]{devrim_springer_2014}%
  \BibitemOpen
  \bibfield  {author} {\bibinfo {author} {\bibfnamefont {A.~D.}\ \bibnamefont {Guclu}}, \bibinfo {author} {\bibfnamefont {P.}~\bibnamefont {Potasz}}, \bibinfo {author} {\bibfnamefont {M.}~\bibnamefont {Korkusinski}},\ and\ \bibinfo {author} {\bibfnamefont {P.}~\bibnamefont {Hawrylak}},\ }\href@noop {} {\emph {\bibinfo {title} {Graphene Quantum Dots}}}\ (\bibinfo  {publisher} {Springer Berlin, Heidelberg},\ \bibinfo {year} {2014})\BibitemShut {NoStop}%
\bibitem [{\citenamefont {Mishra}\ \emph {et~al.}(2021)\citenamefont {Mishra}, \citenamefont {Catarina}, \citenamefont {Wu}, \citenamefont {Ortiz}, \citenamefont {Jacob}, \citenamefont {Eimre}, \citenamefont {Ma}, \citenamefont {Pignedoli}, \citenamefont {Feng}, \citenamefont {Ruffieux}, \citenamefont {Fern{\'a}ndez-Rossier},\ and\ \citenamefont {Fasel}}]{fasel_nature_2021}%
  \BibitemOpen
  \bibfield  {author} {\bibinfo {author} {\bibfnamefont {S.}~\bibnamefont {Mishra}}, \bibinfo {author} {\bibfnamefont {G.}~\bibnamefont {Catarina}}, \bibinfo {author} {\bibfnamefont {F.}~\bibnamefont {Wu}}, \bibinfo {author} {\bibfnamefont {R.}~\bibnamefont {Ortiz}}, \bibinfo {author} {\bibfnamefont {D.}~\bibnamefont {Jacob}}, \bibinfo {author} {\bibfnamefont {K.}~\bibnamefont {Eimre}}, \bibinfo {author} {\bibfnamefont {J.}~\bibnamefont {Ma}}, \bibinfo {author} {\bibfnamefont {C.~A.}\ \bibnamefont {Pignedoli}}, \bibinfo {author} {\bibfnamefont {X.}~\bibnamefont {Feng}}, \bibinfo {author} {\bibfnamefont {P.}~\bibnamefont {Ruffieux}}, \bibinfo {author} {\bibfnamefont {J.}~\bibnamefont {Fern{\'a}ndez-Rossier}},\ and\ \bibinfo {author} {\bibfnamefont {R.}~\bibnamefont {Fasel}},\ }\bibfield  {title} {\bibinfo {title} {Observation of fractional edge excitations in nanographene spin chains},\ }\href {https://doi.org/10.1038/s41586-021-03842-3} {\bibfield  {journal} {\bibinfo  {journal} {Nature}\ }\textbf {\bibinfo
  {volume} {598}},\ \bibinfo {pages} {287} (\bibinfo {year} {2021})}\BibitemShut {NoStop}%
\bibitem [{\citenamefont {Catarina}\ and\ \citenamefont {Fern\'andez-Rossier}(2022)}]{rossier_prb_2022}%
  \BibitemOpen
  \bibfield  {author} {\bibinfo {author} {\bibfnamefont {G.}~\bibnamefont {Catarina}}\ and\ \bibinfo {author} {\bibfnamefont {J.}~\bibnamefont {Fern\'andez-Rossier}},\ }\bibfield  {title} {\bibinfo {title} {Hubbard model for spin-1 haldane chains},\ }\href {https://doi.org/10.1103/PhysRevB.105.L081116} {\bibfield  {journal} {\bibinfo  {journal} {Phys. Rev. B}\ }\textbf {\bibinfo {volume} {105}},\ \bibinfo {pages} {L081116} (\bibinfo {year} {2022})}\BibitemShut {NoStop}%
\bibitem [{\citenamefont {Saleem}\ \emph {et~al.}(2024)\citenamefont {Saleem}, \citenamefont {Steenbock}, \citenamefont {Alhadi}, \citenamefont {Pasek}, \citenamefont {Bester},\ and\ \citenamefont {Potasz}}]{YasserSuperexchangeNanoletter2024}%
  \BibitemOpen
  \bibfield  {author} {\bibinfo {author} {\bibfnamefont {Y.}~\bibnamefont {Saleem}}, \bibinfo {author} {\bibfnamefont {T.}~\bibnamefont {Steenbock}}, \bibinfo {author} {\bibfnamefont {E.~R.~J.}\ \bibnamefont {Alhadi}}, \bibinfo {author} {\bibfnamefont {W.}~\bibnamefont {Pasek}}, \bibinfo {author} {\bibfnamefont {G.}~\bibnamefont {Bester}},\ and\ \bibinfo {author} {\bibfnamefont {P.}~\bibnamefont {Potasz}},\ }\bibfield  {title} {\bibinfo {title} {Superexchange mechanism in coupled triangulenes forming spin-1 chains},\ }\href {https://doi.org/10.1021/acs.nanolett.4c01604} {\bibfield  {journal} {\bibinfo  {journal} {Nano Letters}\ }\textbf {\bibinfo {volume} {24}},\ \bibinfo {pages} {7417} (\bibinfo {year} {2024})},\ \bibinfo {note} {pMID: 38836571},\ \Eprint {https://arxiv.org/abs/https://doi.org/10.1021/acs.nanolett.4c01604} {https://doi.org/10.1021/acs.nanolett.4c01604} \BibitemShut {NoStop}%
\bibitem [{\citenamefont {Baran}\ and\ \citenamefont {Paaske}(2024)}]{Virgilsuperconductor-semiconductorPRB2024}%
  \BibitemOpen
  \bibfield  {author} {\bibinfo {author} {\bibfnamefont {V.~V.}\ \bibnamefont {Baran}}\ and\ \bibinfo {author} {\bibfnamefont {J.}~\bibnamefont {Paaske}},\ }\bibfield  {title} {\bibinfo {title} {Spin-1 haldane chains of superconductor-semiconductor hybrids},\ }\href {https://doi.org/10.1103/PhysRevB.110.064503} {\bibfield  {journal} {\bibinfo  {journal} {Phys. Rev. B}\ }\textbf {\bibinfo {volume} {110}},\ \bibinfo {pages} {064503} (\bibinfo {year} {2024})}\BibitemShut {NoStop}%
\bibitem [{\citenamefont {Manalo}\ \emph {et~al.}(2024)\citenamefont {Manalo}, \citenamefont {Miravet},\ and\ \citenamefont {Hawrylak}}]{ManaloPRBSpin1chain}%
  \BibitemOpen
  \bibfield  {author} {\bibinfo {author} {\bibfnamefont {J.}~\bibnamefont {Manalo}}, \bibinfo {author} {\bibfnamefont {D.}~\bibnamefont {Miravet}},\ and\ \bibinfo {author} {\bibfnamefont {P.}~\bibnamefont {Hawrylak}},\ }\bibfield  {title} {\bibinfo {title} {Microscopic design of a synthetic spin-1 chain in an {InAsP} quantum dot array},\ }\href {https://doi.org/10.1103/PhysRevB.109.085112} {\bibfield  {journal} {\bibinfo  {journal} {Phys. Rev. B}\ }\textbf {\bibinfo {volume} {109}},\ \bibinfo {pages} {085112} (\bibinfo {year} {2024})}\BibitemShut {NoStop}%
\bibitem [{\citenamefont {Pereira}\ \emph {et~al.}(2007)\citenamefont {Pereira}, \citenamefont {Vasilopoulos},\ and\ \citenamefont {Peeters}}]{pereira2007tunable}%
  \BibitemOpen
  \bibfield  {author} {\bibinfo {author} {\bibfnamefont {J.~M.}\ \bibnamefont {Pereira}}, \bibinfo {author} {\bibfnamefont {P.}~\bibnamefont {Vasilopoulos}},\ and\ \bibinfo {author} {\bibfnamefont {F.}~\bibnamefont {Peeters}},\ }\bibfield  {title} {\bibinfo {title} {Tunable quantum dots in bilayer graphene},\ }\href {https://doi.org/10.1021/nl062967s} {\bibfield  {journal} {\bibinfo  {journal} {Nano letters}\ }\textbf {\bibinfo {volume} {7}},\ \bibinfo {pages} {946} (\bibinfo {year} {2007})}\BibitemShut {NoStop}%
\bibitem [{\citenamefont {Recher}\ \emph {et~al.}(2009)\citenamefont {Recher}, \citenamefont {Nilsson}, \citenamefont {Burkard},\ and\ \citenamefont {Trauzettel}}]{recherGrapheneQD2009}%
  \BibitemOpen
  \bibfield  {author} {\bibinfo {author} {\bibfnamefont {P.}~\bibnamefont {Recher}}, \bibinfo {author} {\bibfnamefont {J.}~\bibnamefont {Nilsson}}, \bibinfo {author} {\bibfnamefont {G.}~\bibnamefont {Burkard}},\ and\ \bibinfo {author} {\bibfnamefont {B.}~\bibnamefont {Trauzettel}},\ }\bibfield  {title} {\bibinfo {title} {Bound states and magnetic field induced valley splitting in gate-tunable graphene quantum dots},\ }\href {https://doi.org/10.1103/PhysRevB.79.085407} {\bibfield  {journal} {\bibinfo  {journal} {Physical Review B}\ }\textbf {\bibinfo {volume} {79}},\ \bibinfo {pages} {085407} (\bibinfo {year} {2009})}\BibitemShut {NoStop}%
\bibitem [{\citenamefont {M\"oller}\ \emph {et~al.}(2021)\citenamefont {M\"oller}, \citenamefont {Banszerus}, \citenamefont {Knothe}, \citenamefont {Steiner}, \citenamefont {Icking}, \citenamefont {Trellenkamp}, \citenamefont {Lentz}, \citenamefont {Watanabe}, \citenamefont {Taniguchi}, \citenamefont {Glazman}, \citenamefont {Fal'ko}, \citenamefont {Volk},\ and\ \citenamefont {Stampfer}}]{MollerPRL2021}%
  \BibitemOpen
  \bibfield  {author} {\bibinfo {author} {\bibfnamefont {S.}~\bibnamefont {M\"oller}}, \bibinfo {author} {\bibfnamefont {L.}~\bibnamefont {Banszerus}}, \bibinfo {author} {\bibfnamefont {A.}~\bibnamefont {Knothe}}, \bibinfo {author} {\bibfnamefont {C.}~\bibnamefont {Steiner}}, \bibinfo {author} {\bibfnamefont {E.}~\bibnamefont {Icking}}, \bibinfo {author} {\bibfnamefont {S.}~\bibnamefont {Trellenkamp}}, \bibinfo {author} {\bibfnamefont {F.}~\bibnamefont {Lentz}}, \bibinfo {author} {\bibfnamefont {K.}~\bibnamefont {Watanabe}}, \bibinfo {author} {\bibfnamefont {T.}~\bibnamefont {Taniguchi}}, \bibinfo {author} {\bibfnamefont {L.~I.}\ \bibnamefont {Glazman}}, \bibinfo {author} {\bibfnamefont {V.~I.}\ \bibnamefont {Fal'ko}}, \bibinfo {author} {\bibfnamefont {C.}~\bibnamefont {Volk}},\ and\ \bibinfo {author} {\bibfnamefont {C.}~\bibnamefont {Stampfer}},\ }\bibfield  {title} {\bibinfo {title} {Probing two-electron multiplets in bilayer graphene quantum dots},\ }\href {https://doi.org/10.1103/PhysRevLett.127.256802}
  {\bibfield  {journal} {\bibinfo  {journal} {Phys. Rev. Lett.}\ }\textbf {\bibinfo {volume} {127}},\ \bibinfo {pages} {256802} (\bibinfo {year} {2021})}\BibitemShut {NoStop}%
\bibitem [{\citenamefont {Korkusinski}\ \emph {et~al.}(2023)\citenamefont {Korkusinski}, \citenamefont {Saleem}, \citenamefont {Dusko}, \citenamefont {Miravet},\ and\ \citenamefont {Hawrylak}}]{korkusinski2023spontaneous}%
  \BibitemOpen
  \bibfield  {author} {\bibinfo {author} {\bibfnamefont {M.}~\bibnamefont {Korkusinski}}, \bibinfo {author} {\bibfnamefont {Y.}~\bibnamefont {Saleem}}, \bibinfo {author} {\bibfnamefont {A.}~\bibnamefont {Dusko}}, \bibinfo {author} {\bibfnamefont {D.}~\bibnamefont {Miravet}},\ and\ \bibinfo {author} {\bibfnamefont {P.}~\bibnamefont {Hawrylak}},\ }\bibfield  {title} {\bibinfo {title} {Spontaneous spin and valley symmetry-broken states of interacting massive dirac fermions in a bilayer graphene quantum dot},\ }\href {https://doi.org/10.1021/acs.nanolett.3c02073} {\bibfield  {journal} {\bibinfo  {journal} {Nano Letters}\ }\textbf {\bibinfo {volume} {23}},\ \bibinfo {pages} {7546} (\bibinfo {year} {2023})}\BibitemShut {NoStop}%
\bibitem [{\citenamefont {Albert}\ \emph {et~al.}(2024)\citenamefont {Albert}, \citenamefont {Miravet}, \citenamefont {Saleem}, \citenamefont {Sadecka}, \citenamefont {Korkusinski}, \citenamefont {Bester},\ and\ \citenamefont {Hawrylak}}]{Matthew_PRB_TG_2024}%
  \BibitemOpen
  \bibfield  {author} {\bibinfo {author} {\bibfnamefont {M.}~\bibnamefont {Albert}}, \bibinfo {author} {\bibfnamefont {D.}~\bibnamefont {Miravet}}, \bibinfo {author} {\bibfnamefont {Y.}~\bibnamefont {Saleem}}, \bibinfo {author} {\bibfnamefont {K.}~\bibnamefont {Sadecka}}, \bibinfo {author} {\bibfnamefont {M.}~\bibnamefont {Korkusinski}}, \bibinfo {author} {\bibfnamefont {G.}~\bibnamefont {Bester}},\ and\ \bibinfo {author} {\bibfnamefont {P.}~\bibnamefont {Hawrylak}},\ }\bibfield  {title} {\bibinfo {title} {Optical properties of gated bilayer graphene quantum dots with trigonal warping},\ }\href {https://doi.org/10.1103/PhysRevB.110.155421} {\bibfield  {journal} {\bibinfo  {journal} {Phys. Rev. B}\ }\textbf {\bibinfo {volume} {110}},\ \bibinfo {pages} {155421} (\bibinfo {year} {2024})}\BibitemShut {NoStop}%
\bibitem [{\citenamefont {Miravet}\ \emph {et~al.}(2023)\citenamefont {Miravet}, \citenamefont {Alt\ifmmode \imath \else \i \fi{}nta\ifmmode~\mbox{\c{s}}\else \c{s}\fi{}}, \citenamefont {Rodrigues}, \citenamefont {Bieniek}, \citenamefont {Korkusinski},\ and\ \citenamefont {Hawrylak}}]{Miravet_PRB_QD_Holes}%
  \BibitemOpen
  \bibfield  {author} {\bibinfo {author} {\bibfnamefont {D.}~\bibnamefont {Miravet}}, \bibinfo {author} {\bibfnamefont {A.}~\bibnamefont {Alt\ifmmode \imath \else \i \fi{}nta\ifmmode~\mbox{\c{s}}\else \c{s}\fi{}}}, \bibinfo {author} {\bibfnamefont {A.~W.}\ \bibnamefont {Rodrigues}}, \bibinfo {author} {\bibfnamefont {M.}~\bibnamefont {Bieniek}}, \bibinfo {author} {\bibfnamefont {M.}~\bibnamefont {Korkusinski}},\ and\ \bibinfo {author} {\bibfnamefont {P.}~\bibnamefont {Hawrylak}},\ }\bibfield  {title} {\bibinfo {title} {Interacting holes in gated {W}{$\text{Se}_{2}$} quantum dots},\ }\href {https://doi.org/10.1103/PhysRevB.108.195407} {\bibfield  {journal} {\bibinfo  {journal} {Phys. Rev. B}\ }\textbf {\bibinfo {volume} {108}},\ \bibinfo {pages} {195407} (\bibinfo {year} {2023})}\BibitemShut {NoStop}%
\bibitem [{\citenamefont {\ifmmode~\dot{Z}\else \.{Z}\fi{}ebrowski}\ \emph {et~al.}(2017)\citenamefont {\ifmmode~\dot{Z}\else \.{Z}\fi{}ebrowski}, \citenamefont {Peeters},\ and\ \citenamefont {Szafran}}]{dQDPeeters2017}%
  \BibitemOpen
  \bibfield  {author} {\bibinfo {author} {\bibfnamefont {D.~P.}\ \bibnamefont {\ifmmode~\dot{Z}\else \.{Z}\fi{}ebrowski}}, \bibinfo {author} {\bibfnamefont {F.~M.}\ \bibnamefont {Peeters}},\ and\ \bibinfo {author} {\bibfnamefont {B.}~\bibnamefont {Szafran}},\ }\bibfield  {title} {\bibinfo {title} {Double quantum dots defined in bilayer graphene},\ }\href {https://doi.org/10.1103/PhysRevB.96.035434} {\bibfield  {journal} {\bibinfo  {journal} {Phys. Rev. B}\ }\textbf {\bibinfo {volume} {96}},\ \bibinfo {pages} {035434} (\bibinfo {year} {2017})}\BibitemShut {NoStop}%
\bibitem [{\citenamefont {Paw\l{}owski}\ \emph {et~al.}(2021)\citenamefont {Paw\l{}owski}, \citenamefont {Bieniek},\ and\ \citenamefont {Wo\ifmmode~\acute{z}\else \'{z}\fi{}niak}}]{PawlowskiPRA2021}%
  \BibitemOpen
  \bibfield  {author} {\bibinfo {author} {\bibfnamefont {J.}~\bibnamefont {Paw\l{}owski}}, \bibinfo {author} {\bibfnamefont {M.}~\bibnamefont {Bieniek}},\ and\ \bibinfo {author} {\bibfnamefont {T.}~\bibnamefont {Wo\ifmmode~\acute{z}\else \'{z}\fi{}niak}},\ }\bibfield  {title} {\bibinfo {title} {Valley two-qubit system in a {Mo}{$\text{S}_{2}$}-monolayer gated double quantum dot},\ }\href {https://doi.org/10.1103/PhysRevApplied.15.054025} {\bibfield  {journal} {\bibinfo  {journal} {Phys. Rev. Appl.}\ }\textbf {\bibinfo {volume} {15}},\ \bibinfo {pages} {054025} (\bibinfo {year} {2021})}\BibitemShut {NoStop}%
\bibitem [{\citenamefont {Knothe}\ and\ \citenamefont {Burkard}(2024)}]{dQDKnothePRB2024}%
  \BibitemOpen
  \bibfield  {author} {\bibinfo {author} {\bibfnamefont {A.}~\bibnamefont {Knothe}}\ and\ \bibinfo {author} {\bibfnamefont {G.}~\bibnamefont {Burkard}},\ }\bibfield  {title} {\bibinfo {title} {Extended hubbard model describing small multidot arrays in bilayer graphene},\ }\href {https://doi.org/10.1103/PhysRevB.109.245401} {\bibfield  {journal} {\bibinfo  {journal} {Phys. Rev. B}\ }\textbf {\bibinfo {volume} {109}},\ \bibinfo {pages} {245401} (\bibinfo {year} {2024})}\BibitemShut {NoStop}%
\bibitem [{\citenamefont {Soni}\ \emph {et~al.}(2022)\citenamefont {Soni}, \citenamefont {Kaushal}, \citenamefont {{\c S}en}, \citenamefont {Reboredo}, \citenamefont {Moreo},\ and\ \citenamefont {Dagotto}}]{soniBubbard2BLBQ2022}%
  \BibitemOpen
  \bibfield  {author} {\bibinfo {author} {\bibfnamefont {R.}~\bibnamefont {Soni}}, \bibinfo {author} {\bibfnamefont {N.}~\bibnamefont {Kaushal}}, \bibinfo {author} {\bibfnamefont {C.}~\bibnamefont {{\c S}en}}, \bibinfo {author} {\bibfnamefont {F.~A.}\ \bibnamefont {Reboredo}}, \bibinfo {author} {\bibfnamefont {A.}~\bibnamefont {Moreo}},\ and\ \bibinfo {author} {\bibfnamefont {E.}~\bibnamefont {Dagotto}},\ }\bibfield  {title} {\bibinfo {title} {Estimation of biquadratic and bicubic {{Heisenberg}} effective couplings from multiorbital {{Hubbard}} models},\ }\href {https://doi.org/10.1088/1367-2630/ac7b9c} {\bibfield  {journal} {\bibinfo  {journal} {New Journal of Physics}\ }\textbf {\bibinfo {volume} {24}},\ \bibinfo {pages} {073014} (\bibinfo {year} {2022})}\BibitemShut {NoStop}%
\bibitem [{\citenamefont {Tanaka}\ \emph {et~al.}(2018)\citenamefont {Tanaka}, \citenamefont {Yokoyama},\ and\ \citenamefont {Hotta}}]{tanakaOriginBiquadraticExchange2018}%
  \BibitemOpen
  \bibfield  {author} {\bibinfo {author} {\bibfnamefont {K.}~\bibnamefont {Tanaka}}, \bibinfo {author} {\bibfnamefont {Y.}~\bibnamefont {Yokoyama}},\ and\ \bibinfo {author} {\bibfnamefont {C.}~\bibnamefont {Hotta}},\ }\bibfield  {title} {\bibinfo {title} {Origin of {{Biquadratic Exchange Interactions}} in a {{Mott Insulator}} as a {{Driving Force}} of {{Spin Nematic Order}}},\ }\href {https://doi.org/10.7566/JPSJ.87.023702} {\bibfield  {journal} {\bibinfo  {journal} {Journal of the Physical Society of Japan}\ }\textbf {\bibinfo {volume} {87}},\ \bibinfo {pages} {023702} (\bibinfo {year} {2018})}\BibitemShut {NoStop}%
\bibitem [{\citenamefont {Schollwöck}(2011)}]{schollwock_annals_of_physics_2011}%
  \BibitemOpen
  \bibfield  {author} {\bibinfo {author} {\bibfnamefont {U.}~\bibnamefont {Schollwöck}},\ }\bibfield  {title} {\bibinfo {title} {The density-matrix renormalization group in the age of matrix product states},\ }\href {https://doi.org/https://doi.org/10.1016/j.aop.2010.09.012} {\bibfield  {journal} {\bibinfo  {journal} {Annals of Physics}\ }\textbf {\bibinfo {volume} {326}},\ \bibinfo {pages} {96} (\bibinfo {year} {2011})},\ \bibinfo {note} {january 2011 Special Issue}\BibitemShut {NoStop}%
\bibitem [{\citenamefont {Schollw\"ock}(2005)}]{schollwock_rev_mod_phys_2005}%
  \BibitemOpen
  \bibfield  {author} {\bibinfo {author} {\bibfnamefont {U.}~\bibnamefont {Schollw\"ock}},\ }\bibfield  {title} {\bibinfo {title} {The density-matrix renormalization group},\ }\href {https://doi.org/10.1103/RevModPhys.77.259} {\bibfield  {journal} {\bibinfo  {journal} {Rev. Mod. Phys.}\ }\textbf {\bibinfo {volume} {77}},\ \bibinfo {pages} {259} (\bibinfo {year} {2005})}\BibitemShut {NoStop}%
\bibitem [{\citenamefont {White}(1992)}]{white_prl_1992}%
  \BibitemOpen
  \bibfield  {author} {\bibinfo {author} {\bibfnamefont {S.~R.}\ \bibnamefont {White}},\ }\bibfield  {title} {\bibinfo {title} {Density matrix formulation for quantum renormalization groups},\ }\href {https://doi.org/10.1103/PhysRevLett.69.2863} {\bibfield  {journal} {\bibinfo  {journal} {Phys. Rev. Lett.}\ }\textbf {\bibinfo {volume} {69}},\ \bibinfo {pages} {2863} (\bibinfo {year} {1992})}\BibitemShut {NoStop}%
\bibitem [{\citenamefont {Rakov}\ and\ \citenamefont {Weyrauch}(2022)}]{RakovBLBQHaldanePRB2022}%
  \BibitemOpen
  \bibfield  {author} {\bibinfo {author} {\bibfnamefont {M.~V.}\ \bibnamefont {Rakov}}\ and\ \bibinfo {author} {\bibfnamefont {M.}~\bibnamefont {Weyrauch}},\ }\bibfield  {title} {\bibinfo {title} {Bilinear-biquadratic spin-1 model in the haldane and dimerized phases},\ }\href {https://doi.org/10.1103/PhysRevB.105.024424} {\bibfield  {journal} {\bibinfo  {journal} {Phys. Rev. B}\ }\textbf {\bibinfo {volume} {105}},\ \bibinfo {pages} {024424} (\bibinfo {year} {2022})}\BibitemShut {NoStop}%
\bibitem [{\citenamefont {Sadecka}\ \emph {et~al.}(2024)\citenamefont {Sadecka}, \citenamefont {Saleem}, \citenamefont {Miravet}, \citenamefont {Albert}, \citenamefont {Korkusinski}, \citenamefont {Bester},\ and\ \citenamefont {Hawrylak}}]{sadecka2023electrically}%
  \BibitemOpen
  \bibfield  {author} {\bibinfo {author} {\bibfnamefont {K.}~\bibnamefont {Sadecka}}, \bibinfo {author} {\bibfnamefont {Y.}~\bibnamefont {Saleem}}, \bibinfo {author} {\bibfnamefont {D.}~\bibnamefont {Miravet}}, \bibinfo {author} {\bibfnamefont {M.}~\bibnamefont {Albert}}, \bibinfo {author} {\bibfnamefont {M.}~\bibnamefont {Korkusinski}}, \bibinfo {author} {\bibfnamefont {G.}~\bibnamefont {Bester}},\ and\ \bibinfo {author} {\bibfnamefont {P.}~\bibnamefont {Hawrylak}},\ }\bibfield  {title} {\bibinfo {title} {Electrically tunable fine structure of negatively charged excitons in gated bilayer graphene quantum dots},\ }\href {https://doi.org/10.1103/PhysRevB.109.085434} {\bibfield  {journal} {\bibinfo  {journal} {Physical Review B}\ }\textbf {\bibinfo {volume} {109}},\ \bibinfo {pages} {085434} (\bibinfo {year} {2024})}\BibitemShut {NoStop}%
\bibitem [{\citenamefont {{Digital Research Alliance of Canada}}(2025)}]{alliancecan2025}%
  \BibitemOpen
  \bibfield  {author} {\bibinfo {author} {\bibnamefont {{Digital Research Alliance of Canada}}},\ }\href@noop {} {\bibinfo {title} {{Digital Research Alliance of Canada}}},\ \bibinfo {howpublished} {\url{https://alliancecan.ca}} (\bibinfo {year} {2025}),\ \bibinfo {note} {accessed: 2025-07-26}\BibitemShut {NoStop}%
\bibitem [{\citenamefont {Saleem}\ \emph {et~al.}(2023)\citenamefont {Saleem}, \citenamefont {Sadecka}, \citenamefont {Korkusinski}, \citenamefont {Miravet}, \citenamefont {Dusko},\ and\ \citenamefont {Hawrylak}}]{saleem2023theory}%
  \BibitemOpen
  \bibfield  {author} {\bibinfo {author} {\bibfnamefont {Y.}~\bibnamefont {Saleem}}, \bibinfo {author} {\bibfnamefont {K.}~\bibnamefont {Sadecka}}, \bibinfo {author} {\bibfnamefont {M.}~\bibnamefont {Korkusinski}}, \bibinfo {author} {\bibfnamefont {D.}~\bibnamefont {Miravet}}, \bibinfo {author} {\bibfnamefont {A.}~\bibnamefont {Dusko}},\ and\ \bibinfo {author} {\bibfnamefont {P.}~\bibnamefont {Hawrylak}},\ }\bibfield  {title} {\bibinfo {title} {Theory of excitons in gated bilayer graphene quantum dots},\ }\href {https://doi.org/10.1021/acs.nanolett.3c00406} {\bibfield  {journal} {\bibinfo  {journal} {Nano Letters}\ }\textbf {\bibinfo {volume} {23}},\ \bibinfo {pages} {2998} (\bibinfo {year} {2023})}\BibitemShut {NoStop}%
\end{thebibliography}%

\clearpage
\onecolumngrid
\appendix
\section*{Supplemental Material}

\setcounter{equation}{0}
\setcounter{figure}{0}
\setcounter{table}{0}

\renewcommand{\theequation}{S\arabic{equation}}
\renewcommand{\thefigure}{S\arabic{figure}}
\renewcommand{\thetable}{S\arabic{table}}


\section{Details of the Model}
\subsection{Tight-Binding model}
\label{sect:TB model}

We consider a Bernal-stacked bilayer graphene system as illustrated in Fig. 1(a) in the main text, where the bottom (top) layer comprises sublattices $A_1$ and $B_1$ ($A_2$ and $B_2$). The in-plane nearest-neighbor (NN) bond length is $a = 0.143$ nm, and the interlayer separation is $h = 0.335$ nm. The unit cell vectors are defined as $\vec{a}_1 = a(0, \sqrt{3})$ and $\vec{a}_2 = \frac{a}{2}(3, -\sqrt{3})$. We model the BLG using a rhomboidal computational domain with  $N_1 = N_2 = 901$ unit cells along $\vec{a}_1$ and $\vec{a}_2$, containing over three million carbon atoms.

To mitigate finite-size effects, we impose periodic boundary conditions, which yield a discrete set of $k$-points in momentum space \cite{Miravet_PRB_QD_Holes}. Using Bloch's theorem, the wavefunction on sublattice $l$ takes the form $\ket{\phi^l_{\vec{k}}} = \frac{1}{\sqrt{N_{\rm UC}}} \sum_{\vec{R}_l} e^{i \vec{k} \cdot \vec{R}_l} \ket{\vec{R}_l}$, where $N_{\rm UC} = N_1 \times N_2$ is the total number of unit cells and $\ket{\vec{R}_l}$ denotes a $p_z$ orbital at position $\vec{R}_l$.

To open a gap in the bulk band structure, we introduce an electric potential difference between the two layers of the BLG, creating a downward electric field, which increases the offset energy of the electrons in the bottom layer by $V_E$, as shown in Fig. 1(b) of the main text.
The bulk Hamiltonian in the sublattice basis ($A_1$, $B_1$, $A_2$, $B_2$) then reads 
\begin{equation}\label{eq:BulkHamiltonian}
H_{\textrm{bulk}}(\vec{k}) = 
    \begin{pmatrix}
        \frac{V_E}{2} & \gamma_0 f(\vec{k}) & 0 & 0\\
         \gamma_0 f^*(\vec{k})  & \frac{V_E}{2} & \gamma_1 &
        0 \\ 0 & \gamma_1 &  -\frac{V_E}{2} &  \gamma_0 f(\vec{k}) \\
        0 & 0 & \gamma_0 f^*(\Vec{k}) & -\frac{V_E}{2} \\
    \end{pmatrix},
\end{equation}
where $\gamma_0 = -2.5$ eV is the NN intralayer hopping, and $\gamma_1 = 0.34$ eV corresponds to the dominant interlayer coupling between $B_1$ and $A_2$ \cite{sadecka2023electrically,saleem2023theory}. 

The form factor $f(\vec{k}) = \sum_{j=1}^3 e^{i\vec{k} \cdot \vec{\delta}_j}$ involves the three in-plane NN vectors, defined as $\vec{\delta}_1 = \vec{b}$, $\vec{\delta}_2 = \vec{b} - \vec{a}_2 - \vec{a}_1$, and $\vec{\delta}_3 = \vec{b} - \vec{a}_1$, where $\vec{b} = \frac{a}{2}(1, \sqrt{3})$.

Diagonalizing $H_{\textrm{bulk}}(\vec{k})$ for a given $\vec{k}$ yields eigenstates $\ket{\psi^p_{\vec{k}}} = \sum_l A^p_{\vec{k}, l} \ket{\phi^l_{\vec{k}}}$, where $p$ indexes the four energy bands, and $l$ runs over the sublattices.

\subsection{Gate-defined Quantum Dots} 
\label{sect:GateQD}

We can apply lateral confining potentials to each layer, effectively confining electrons to the center of the system~\cite{Miravet_PRB_QD_Holes}. That potential can be modelled by a Gaussian potential given by: 
\begin{equation}\label{eq:ConfiningPotential_dot}
    {V}_{\rm{QD}}(x,y) =- V_0 \, e^{-\frac{|\vec{\rho}|^2}{R_{\rm{QD}}^2}}
\end{equation}
where $\vec{\rho} = (x,y)$ represents the continuous coordinate on the $xy$ plane, $R_{QD}$ is the QD radius and $V_0$ represents the depth of the confining potential. 

We can also create a double quantum dot system by applying a lateral confining potential in two different areas of space. The potential can be represented as a sum of two Gaussian functions centered at two different points in space
\begin{equation}\label{eq:ConfiningPotential_double_dot}
    {V}_{\rm{dQD}}(x,y) = {V}_{QD}(x,y-d/2) + {V}_{QD}(x,y+d/2) .
\end{equation}
Here, $d$ is the distance between the two QD centers. 

We can write the Hamiltonian of the gated confined electrons as
\begin{equation}\label{eq:QDHamiltonian}
    \hat{H}_{\rm{QD}} = \hat{H}_{\textrm{bulk}} + \hat{V}_{\rm{ext}}.
\end{equation}
Where $V_{\rm{ext}}$ can be the single QD (${V}_{QD}$) or the double QD potential (${V}_{dQD}$). 
We can obtain the single-particle eigenstates of the confined system by expanding them in the basis of band eigenstates of the bulk Hamiltonian in Eq.~(\ref{eq:BulkHamiltonian}),
\begin{equation}
    \varphi_s(\vec{r})= \sum_{\vec{k}}\sum_{p} B^s_{p,\vec{k}} \, \psi^p_{\vec{k}}(\vec{r}),
\end{equation}
where $\vec{k}$ represents the discrete set of points in the momentum space obtained after imposing periodic boundary conditions, and the index $p$ runs over the bulk bands. Substituting this expression into Schrodinger's equation, we obtain the eigenvalue problem
\begin{equation}\label{eq:Schrodinger}
    \varepsilon_{p,\vec{k}}\, B^s_{p,\vec{k}} + \sum_{p', \vec{k}'}\bra{\psi_{\vec{k}}^p} \hat{V}_{\rm{ext}} \ket{\psi_{\vec{k}'}^{p'}} B^s_{p',\vec{k}'} = E_s B^s_{p,\vec{k}},
\end{equation}
where $\varepsilon_{p,\vec{k}}$ are the band energies of BLG. By solving Eq.~(\ref{eq:Schrodinger}), we can obtain the energies $E_s$ of a confined electron and construct its wave function using the coefficients $B^s_{p,\vec{k}}$. Note that the confining potential couples wave-vectors $\vec{k}$ and $\vec{k}'$ on bands $p$ and $p'$ via the matrix element
\begin{equation}
\bra{\psi_{\vec{k}}^p}\hat{V}_{\rm{ext}}\ket{\psi_{\vec{k}'}^{p'}} = V_{\vec{k},\vec{k}'}\sum_{l} (A^{p}_{\vec{k},l})^*\, A^{p'}_{\vec{k}',l} e^{i\left(\vec{k}'-\vec{k}\right)\cdot \vec{d}_l} ,
\end{equation}
Here $\vec{d}_l$ is the relative position of an atom on sublattice $l$ in a unit cell, and $V_{\vec{k},\vec{k}'}$ is the Fourier transform of the confining potential as defined in Eqs.~(\ref{eq:ConfiningPotential_dot}) and~(\ref{eq:ConfiningPotential_double_dot}). For the single QD, the explicit expression for $V_{\vec{k},\vec{k}'}$ is given by

\begin{equation}\label{eq:FourierTransform}
    V_{\vec{k},\vec{k}'} =  V_0 \frac{A_k}{4\pi} R_{\rm{QD}}^2 e^{-\frac{R_{\rm{QD}}^2 |\vec{k}-\vec{k}'|^2}{4}},
\end{equation}
where $A_k = \frac{8\pi^2}{3\sqrt{3}a^2N_{\rm UC}}$ represents the reciprocal lattice unit cell area. Note that the decay rate on the reciprocal space is proportional to the QD radius, that is, as the QD size increases, the coupling between different $\vec{k}$ decreases.
In the case of the double QD, the expression for $V_{\vec{k},\vec{k}'}$ is given by

\begin{equation}\label{eq:FourierTransformdQD}
    V_{\vec{k},\vec{k}'} =  V_0 \frac{A_k}{4\pi} R_{\rm{QD}}^2 e^{-\frac{R_{\rm{QD}}^2 |\vec{k}-\vec{k}'|^2}{4}} \cos((k_y-k'_y)\frac{d}{2}),
\end{equation}
Compared to the single quantum dot, in this case, we have an extra oscillatory term, modulated by an exponential decay.

Since we are interested in the low-energy spectrum of the QDs, only $\vec{k}$ states near the Fermi level contribute. We impose an energy cut-off of $E_{\rm cut}=600$ meV on our band states $\ket{\psi_{\vec{k}}^p}$, including only those band states with energy $\varepsilon_{p,\vec{k}}$ in the range $|\varepsilon_{p,\vec{k}}| \leq E_{\rm cut}$. This ensures the convergence of low-energy single-particle QD states, which are composed of $\vec{k}$ vectors around the $\pm K$ points. 


\section{Single QD}
\label{sect:sigleQD}

\begin{figure}[ht]
	a)\includegraphics[width=12cm]{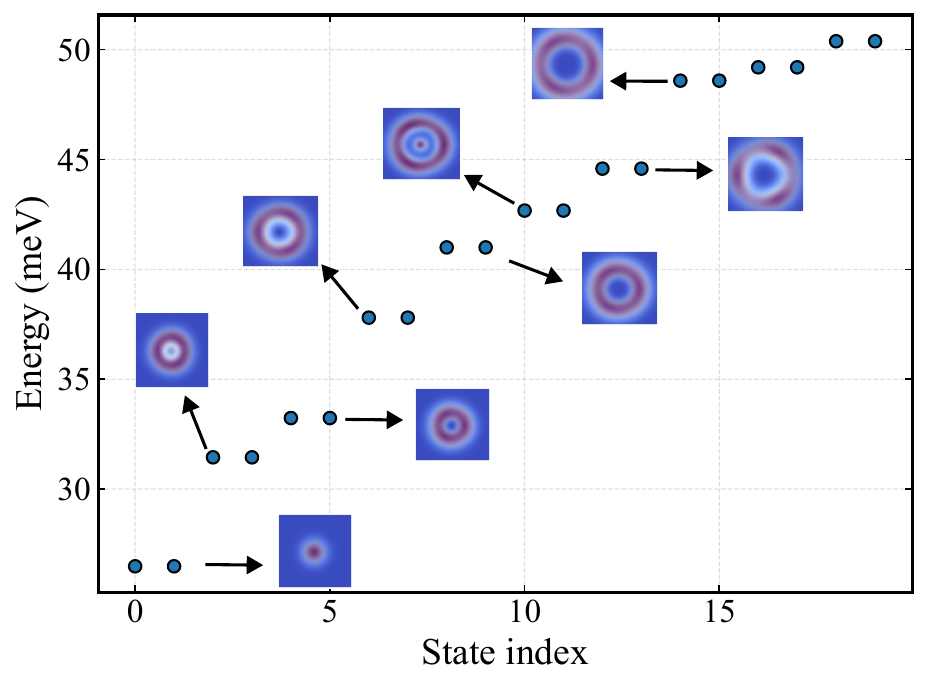}
    b)\includegraphics[width=12cm]{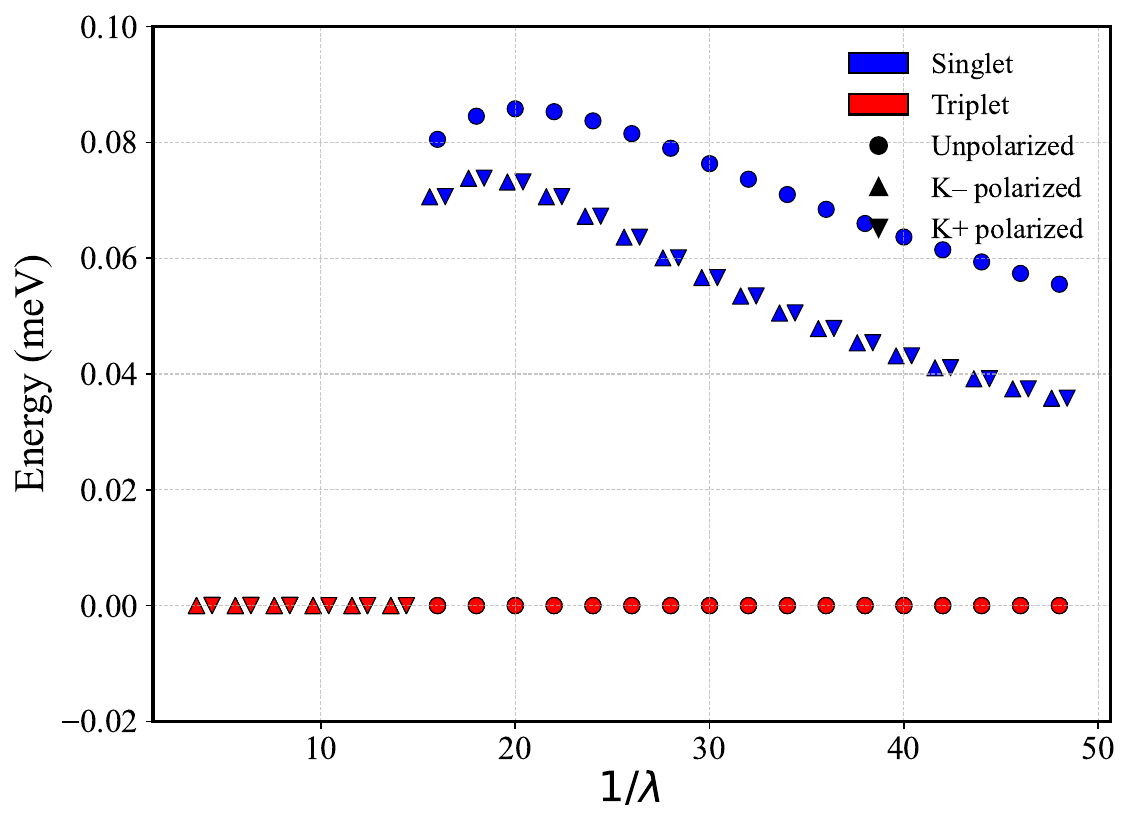}
	\caption{(a) Low-energy levels of a single quantum dot. Each state shown is doubly degenerate by spin; only the valley-degenerate states are explicitly labeled. Insets display the corresponding electron densities for each shell, illustrating the increasing spatial extension of the states with energy.  
    (b) Low-energy spectrum for two electrons as a function of interaction strength. The ground state remains a triplet across the entire range of interaction strengths.
    }

	\label{fig:QD_levels}
\end{figure}

The QD energy spectrum is obtained by solving the eigenvalue problem defined in Eq.~(\ref{eq:Schrodinger}). Figure \ref{fig:QD_levels} shows the resulting energy levels for a displacement field of $V_E = 150$~meV, a QD radius of $R_{QD} = 20$~nm, and a confining potential depth of $V_0 = 50$~meV. The energy levels exhibit valley doublets, which become fourfold degenerate when spin is taken into account. In each valley, the spectrum reveals features characteristic of a two-dimensional harmonic oscillator: a single lowest-energy "s" state, followed by a "p" shell with two states, a "d" shell with three states, and so on. However, in contrast to the ideal harmonic oscillator, the underlying graphene lattice and orbital structure introduce additional splitting within each shell.

After including electron-electron interactions, we compute the ground state of the quantum dot occupied by two electrons. Figure~\ref{fig:QD_levels}b shows how the low-energy spectrum evolves as a function of the interaction strength. For each value of interaction, the ground-state energy is set to zero as the reference. We evaluate the expectation value of the total spin operator $\hat{S}^2$ for each low-energy state, identifying triplet states (shown in red) and singlet states (shown in blue). Notably, the triplet state consistently emerges as the ground state across the full range of interaction strengths. However, the energy gap between the triplet and the first excited singlet state decreases as the Coulomb interaction becomes weaker.

We also classify these states according to their valley polarization, defined as $\nu = N_K - N_{-K}$, where $N_{\tau}$ denotes the number of electrons in valley $\tau$. Since the Hamiltonian conserves $N_{\tau}$, the values of $\nu$ are integers. Valley-polarized states ($\nu \neq 0$) appear as doublets with opposite polarization values $\pm \nu$. In Fig.~\ref{fig:QD_levels}b, we indicate the valley polarization using symbols: a circle denotes valley-unpolarized states, while pointing up and down triangles represent states with more electrons in the $K$ and $-K$ valleys, respectively.

For strong electron-electron interactions, the ground state consists of two degenerate, valley-polarized triplet states, followed by two valley-polarized singlets. As the interaction strength decreases, the ground state transitions to a valley-unpolarized triplet, followed by two valley-polarized singlets and one valley-unpolarized singlet. The last order of the low-energy spectrum has been observed experimentally~\cite{MollerPRL2021}.

 \subsection{Many-body Interactions}
 
 After obtaining the single particle states, we now consider the problem where $N$ electrons occupy the lowest energy states of the system. The many-body Hamiltonian can be written as
\begin{equation}
H=\sum_{s} E_s c^\dagger_s c_s +  \frac{\lambda}{2}\sum_{p,q,r,s} \bra{pq}V\ket{rs} c^\dagger_p c^\dagger_q c_r c_s.
\label{eq:manybodyHam}
\end{equation}
$E_s$  are single-particle energies, obtained  from Eq.~\ref{eq:Schrodinger}. The operator $c^\dagger_s $ ($c_s$) creates (annihilates) an electron in the state $s$. $\bra{pq}V\ket{rs}$ are the Coulomb matrix elements in terms of the single-particle orbitals~ 

\begin{align}
    \bra{pq}V\ket{rs} =&\delta_{\sigma_p,\sigma_s}\delta_{\sigma_q,\sigma_r}\int d\vec{r}_{1}\int d\vec{r}_{2}\varphi_p(\vec{r}_{1})^{\ast}\varphi_{q}(\vec{r}_{2})^{\ast} \nonumber\\
    &\times V(\vec{r}_{2}-\vec{r}_{1})\varphi_{r}(\vec{r}_{2})\varphi_{s}(\vec{r}_{1}),
\end{align}
the Coulomb potential $V(\vec{r}_{2}-\vec{r}_{1})=\frac{e^2}{4\pi\epsilon_0}\frac{1}{|\vec{r}_{2}-\vec{r}_{1}|}$, $e$ is the electron charge, and $\epsilon_0$ is the vacuum permittivity. $\sigma_s$ represent the spin of the given state $s$ and $\delta_{\sigma,\sigma'}$ are Kronecker deltas. The parameter $\lambda\in [0,1]$ is used to tune the ratio between electron-electron interactions and kinetic energy. 

We obtain the low-energy spectrum of the many-body Hamiltonian represented by Eq.~(\ref{eq:manybodyHam}) by using exact diagonalization.

\subsection{Quantum Dot as an Effective spin-1 Site}

The Heisenberg Hamiltonian for a Spin-1 chain is given by 

\begin{equation}
H=J\sum_{i}^{L-1} \vec{S}_i\cdot \vec{S}_{i+1}
\label{eq:manybodyHam_}
\end{equation}
where $\vec{S}_i$ is the  spin operator acting on site $i$, and $J$ is the exchange coupling, and $L$ is the number of sites. For even $L$, the low-energy spectrum consists of a singlet ground state, followed by a triplet and a quintuplet. In the specific case of $L=2$, the Hamiltonian can be rewritten using the identity $\vec{S}_1\cdot \vec{S}_{2}=\frac{1}{2}\left(\vec{S}_{\rm tot}^2-4\right)$, yielding
\begin{equation}
H=\frac{J}{2}\left(\vec{S}_{\rm tot}^2-4\right)
\label{eq:manybodyHam__}
\end{equation}
From this, we can directly compute the energies of the spin multiples
\begin{equation}
E_S = -2J, \quad
E_T = -J, \quad
E_Q = J.
\end{equation}

The energy gap between the triplet and quintuplet is $\Delta_{TQ}=2J$ , which is exactly twice the singlet–triplet gap  $\Delta_{ST}=J$. This fixed ratio imposes a constraint that makes it challenging to accurately match the spectrum of the double quantum dot system using only the Heisenberg model (see Fig.~\ref{fig:BLBQfit}). However, adding a biquadratic term (which also conserves total spin) to the Heisenberg model, we can fit the low-energy spectrum of the double-dot with 4 electrons more accurately.

\begin{figure}[ht]
\centering     
\includegraphics[width=16cm]{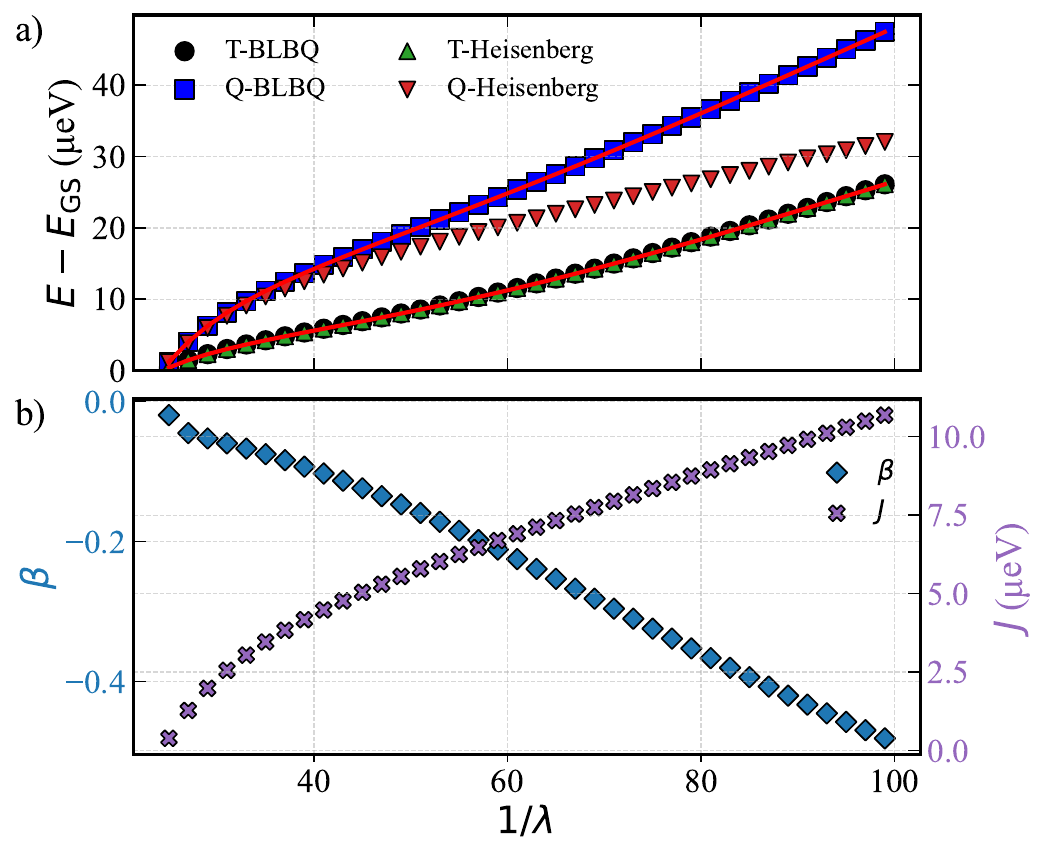}

\caption{
Top panel: Triplet and quintuplet energies obtained from the fitted BLBQ and Heisenberg models as a function of interaction strength. The blue dashed lines show the energies of the double quantum dot. As discussed in the main text, the BLBQ Hamiltonian accurately reproduces the low-energy spectrum of two coupled quantum dots, each containing two electrons.  
Bottom panel: Fitted values of the BLBQ parameters $\beta$ and $J$ as functions of the electron-electron interaction strength.}
  
\label{fig:BLBQfit}
\end{figure}

After adding a biquadratic term, we have the bilinear–biquadratic spin-1 Hamiltonian (BLBQ)
\begin{equation}
H=J\sum_{i}^{L-1} \left( \vec{S}_i\cdot  \vec{S}_{i+1} + \beta \left( \vec{S}_i\cdot  \vec{S}_{i+1} \right)^2 \right)
\label{eq:manybodyHam___}
\end{equation}
where $\beta$ is a tunable parameter that controls the strength of the biquadratic interaction. For two spins, using the same identity as before, the energy levels are
\begin{equation}
E_S = J(-2 + 4\beta), \quad
E_T = J(-1 + \beta), \quad
E_Q = J(1 + \beta)
\end{equation}
 Thus, the gaps are given by
 \begin{equation}
\Delta_{TQ}=2J , \quad \Delta_{ST}=J(1-3\beta).
\end{equation}

\section{Effect of interlayer hopping $\gamma3$ and $\gamma4$}
If we consider more that first nearest interlayer hopping, the Hamiltonian describing bulk BLG is given by \cite{Matthew_PRB_TG_2024}:
\begin{equation}\label{eq:BulkHamiltonian}
H_{\textrm{bulk}}(\Vec{k}) = 
    \begin{pmatrix}
        \frac{V_E}{2} & \gamma_0 f(\Vec{k}) & \gamma_4 f(\Vec{k}) & \gamma_3 f^*(\Vec{k})\\
         \gamma_0 f^*(\Vec{k})  & \frac{V_E}{2} & \gamma_1 &
        \gamma_4 f(\Vec{k}) \\ \gamma_4 f^*(\Vec{k}) & \gamma_1 &  -\frac{V_E}{2} &  \gamma_0 f(\Vec{k}) \\
        \gamma_3 f(\Vec{k}) & \gamma_4 f^*(\Vec{k}) & \gamma_0 f^*(\Vec{k}) & -\frac{V_E}{2} \\
    \end{pmatrix},
\end{equation}
where the parameters $\gamma_3$ and $\gamma_4$ describe the next NN interlayer hopping but differ in that $\gamma_3$ represents the next NN interlayer hopping between pairs of orbitals localized on unstacked atoms, whereas $\gamma_4$ is between an orbital on a stacked and an unstacked atom. The inclusion of $\gamma_3$ introduces the effects of TW, whereas $\gamma_4$ breaks the electron-hole symmetry. We set $\gamma_3 = \gamma_4 = 0.12|\gamma_0|$ \cite{korkusinski2023spontaneous,sadecka2023electrically}.

\begin{figure}[ht]
\centering     
a)\includegraphics[width=\columnwidth]{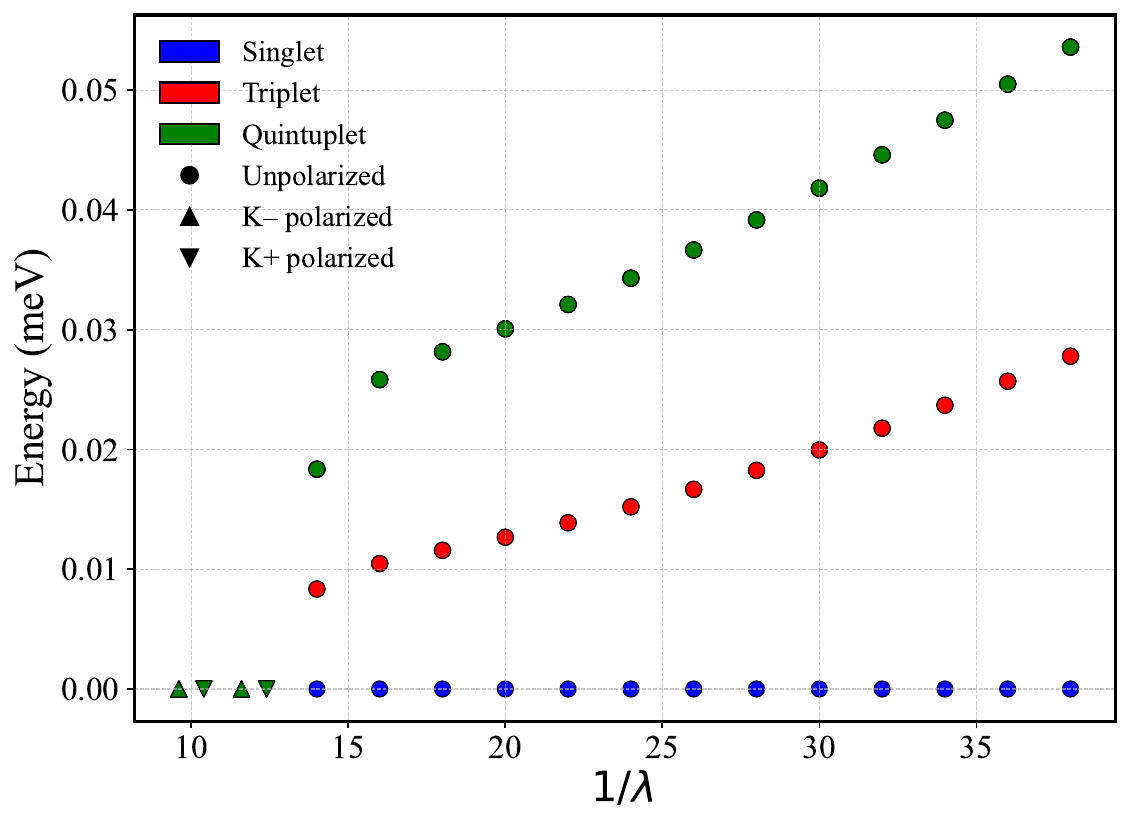}
\caption{(a) Including $\gamma_3 = \gamma_4 = 0.12|\gamma_0|$. Low-energy spectrum for four electrons in a double quantum dot as a function of interaction strength. For weak interactions, the ground state is a valley-unpolarized singlet, while for strong interactions, it becomes a valley-polarized quintuplet. In the regime where the ground state is a singlet, the first and second excited states are a triplet and a quintuplet, respectively, closely resembling the spectrum of a spin-1 Heisenberg antiferromagnetic chain with $L = 2$ sites.}

\label{fig:DQD_energy_levels_TG}
\end{figure}

The low energy spectrum for a double QD after including $\gamma_3$ and $\gamma_4$ is shown in Figure~\ref{fig:DQD_energy_levels_TG}. Note that the structure of the levels does not change quantitatively compared to the case considered in the main text. 

\end{document}